\documentclass[useAMS,usenatbib]{mn2e}
\usepackage[letterpaper, totalwidth=480pt, totalheight=680pt]{geometry}
\usepackage{psfig,ifthen,dcolumn,url}
\usepackage[breaklinks]{hyperref}
\usepackage{hypcap}
\newcommand{\Ttau} {\hbox{$T(\tau)$} relation}
\newcommand{\figdir}{figs}
\newcommand{\mfigur}[3]
    {\begin{figure}
        \centerline{\psfig{figure=\figdir/#1.ps,height=#2}}
        \caption{\label{#1}#3}
    \end{figure}}
\def\aatab#1#2#3#4#5#6{\ifthenelse{\equal{*}{#1}}
{\begin{table*}\caption[]{\label{#2} #3}
  \begin{flushleft}
    \begin{tabular}{#4}
      \hline\noalign{\smallskip} #5
      \noalign{\smallskip} \hline \noalign{\smallskip} #6
      \noalign{\smallskip} \hline
    \end{tabular}
  \end{flushleft}
\end{table*}}
{\begin{table}\scriptsize\caption[]{\label{#1} #2}
  \begin{flushleft}
    \begin{tabular}{#3}
      \hline\noalign{\smallskip} #4
      \noalign{\smallskip} \hline \noalign{\smallskip} #5
      \noalign{\smallskip} \hline
    \end{tabular}
  \end{flushleft}
\end{table}}}
\newcommand{\oh}[1]{\omit\hidewidth #1\hidewidth}
\newcommand{\be}[1]{\begin{equation}\label{#1}}
\newcommand{\ee}{\end{equation}}
\newcommand{\bea}[1]{\begin{eqnarray}\label{#1}}
\newcommand{\eea}{\end{eqnarray}}
\newcommand{\lwig}{{\leavevmode\kern0.3em\raise.3ex\hbox{$<$}
                    \kern-0.8em\lower.7ex \hbox{$\sim$}\kern0.3em}}
\newcommand{\dd} [2]{{{{\rm d}{#1}\over{\rm d}{#2}}}}

\newcommand{\stimes}{\negthinspace\times\negthinspace}
\renewcommand{\d}{{\rm d}}
\newcommand{\dxdy}[2]{\frac{\upartial #1}{\upartial #2}}
\newcommand{\dxdycz}[3]{\left( \frac{ \upartial #1 }{ \upartial #2 }\right)_{#3}}
\newlength{\pwidth}
\newcommand{\ppmm}[2]{\lower-.7ex\hbox{\scriptsize$#1$}\settowidth{\pwidth}%
            {\scriptsize$#2$}\kern-\pwidth\lower.5ex\hbox{\scriptsize$#2$}}
\newcommand{\sfrac}[2]{{\leavevmode
                    \kern.1em\raise.5ex\hbox{\the\scriptfont0 #1}
                    \kern-.13em\raise.25ex\hbox{\the\scriptfont0 /}
                    \kern-0.12em\lower.25ex\hbox{\the\scriptfont0 #2}}}
\newcommand{\eg}  {e.g.}            
\newcommand{\cf}  {cf.}             
\newcommand{\ie}  {i.e.}            
\begin{document}
    \title[Improving stellar models: $\alpha$ calibration]
          {Improvements to stellar structure models, based on a grid of 3D
				convection simulations. II. Calibrating the mixing-length formulation}
    \author[R. Trampedach, et al.]{Regner Trampedach$^{1,2}$\thanks{E-mail: trampeda@lcd.colorado.edu},
     Robert F. Stein$^{3}$,
     J{\o}rgen {Christensen-Dalsgaard}$^{1}$,  \newauthor
     {\AA}ke Nordlund$^{4}$
     and Martin Asplund$^{5}$\\
    $^{1}$Stellar Astrophysics Centre, Department of Physics and Astronomy,
          Ny Munkegade 120, Aarhus University,\\DK--8000 Aarhus C, Denmark\\
    $^{2}$JILA, University of Colorado and National Institute of Standards
          and Technology, 440 UCB, Boulder, CO 80309, USA\\
    $^{3}$Department of Physics and Astronomy, Michigan State University
          East Lansing, MI 48824, USA\\
    $^{4}$Astronomical Observatory/Niels Bohr Institute, Juliane Maries Vej 30,
          DK--2100 Copenhagen {\O}, Denmark\\
    $^{5}$Research School of Astronomy and Astrophysics,
          Mt.\ Stromlo Observatory, Cotter Road, Weston ACT 2611, Australia}

    \date{Received \today / Accepted \today}
    \pagerange{\pageref{firstpage}--\pageref{lastpage}} \pubyear{2013}

    \maketitle

    \label{firstpage}

    \begin{abstract}
        We perform a calibration of the mixing length of convection in stellar
        structure models against realistic 3D radiation-coupled hydrodynamics
        (RHD) simulations of convection in stellar surface layers, determining
        the adiabat deep in convective stellar envelopes.

        The mixing-length
        parameter $\alpha$ is calibrated by matching averages of the
        3D simulations to 1D stellar envelope models, ensuring identical
        atomic physics in the two cases. This is done for a previously published
        grid of solar-metallicity convection simulations, covering from 4\,200\,K to 6\,900\,K
        on the main sequence, and 4\,300--5\,000\,K for giants with
        $\log g=2.2$.

	    Our calibration results in an $\alpha$ varying from 1.6 for the
        warmest dwarf, which is just cool enough to admit a convective
        envelope, and up to 2.05 for the coolest dwarfs in our
        grid. In between these is a triangular plateau of
        $\alpha\sim 1.76$. The Sun is located on this plateau and has seen
        little change during its evolution so far. When stars ascend the
        giant branch, they largely do so along tracks of constant $\alpha$,
        with $\alpha$ decreasing with increasing mass.
    \end{abstract}


\section{Introduction}
\label{sect:intro}
Due to the lack of a better theory of convection in stars,
the mixing-length theory \citep[][MLT]{boehm:mlt} has been in use for more
than half a century.
By far the largest part of the solar convection zone is very close to
adiabatic, and the stratification in the bulk of the convection zone is
therefore determined by the adiabatic 
gradient, $\nabla_{\rm ad} = (\upartial\ln T/\upartial\ln p)_{\rm ad}$, of
temperature, $T$, and pressure, $p$.
Convection is so efficient that the actual gradient, $\nabla$, need only be
slightly larger than the adiabatic one, to transport the entire energy flux.
In most of the convection zone the super-adiabatic gradient, $\nabla_s =
\nabla-\nabla_{\rm ad}$, is small enough, $\nabla_s \lwig 10^{-5}$, to have
no effect on the structure of the star.
We therefore have no need for a theory of convection here,
except as a means to determine the adiabat of the convection zone.

This picture changes dramatically near the boundaries, especially near the
upper boundary of a convective envelope. Here convection 
becomes exceedingly inefficient in transporting the energy flux, 
as radiative energy transport takes over.
For the Sun, this layer of
appreciable superadiabaticity only
occurs in the outermost $\sim$1\,Mm, just below the photosphere---the region
where the gas becomes optically transparent and radiation escapes.
This layer, however thin, is crucial for the star as
a whole, as it is the star's insulation against the cold of space. In other
words, this layer constitutes the outer boundary condition for stellar models.
It is the interaction between convection and radiative transfer in this layer
that sets up the entropy jump at the top of the convection zone, and hence
determines the adiabat of the convection zone.
Modelling stellar atmospheres is complicated and computationally expensive
and beyond the scope of normal stellar structure calculations. Instead, the
structure modeller adopts results from atmosphere models  that describe the
radiative transfer through the atmosphere, and, in particular, the photospheric
transition from optically thick to optically thin.
One solution that we also adopt is to describe this transition by the
behaviour of temperature, $T$, with optical depth, $\tau$, known as a {\Ttau}
\citep[{\eg}][]{henyey:stel-evol3}. With the radiative components fixed, the adiabat
is then determined by the parameters of the MLT formulation.

Hotter, late A-type stars, close to the line in the HR diagram where stars
loose their convective envelopes, have such shallow convective envelopes that
they do not reach an adiabat. There, convection is inefficient and transports
only a minor fraction of the flux, while the temperature fluctuations and the
turbulent pressure are appreciable even at the bottom of the convection zone.
This is the prediction of any realistic
formulation of convection, and is confirmed by convection simulations
\citep{freytag:2D-DA-WD+Astars+Sun,kupka:2D-A-star-atm,freytag:Astars2004,trampedach:Astars2004}.
At the other end of the spectrum we find the cool M dwarfs that are fully
convective. Giants, on the other hand, are dominated by extensive convective
envelopes, and strongly condensed cores. This spans the range
of stars for which the present work applies.

With the advances in atomic physics as applied to astrophysics, {\ie}, the equation of state
\citep[EOS, {\eg},][]{mhd1,saumon:dense-EOS,Qmhd,GDZ:rel-e-MHD,rogers:newOPAL,militzer:HHeEOS-recalPltMR} and opacities
\citep{kur:missolar,rog:OPAL,alexander:FergusonOPAC,OP1,iglesias:OPAL2,OP05,ferguson:GrainOpac},
by far the most uncertain aspects of stellar models are associated with
dynamical phenomena: semi-convection, rotational mixing, mixing by g modes,
convective overshooting and the most prominent: convection itself.
Phenomena that all involve turbulence, non-locality and non-linearity, which
make them less amenable to analytical formulations.

The present paper is part of an effort to improve on stellar structure
models, by using results from a grid of realistic 3D hydrodynamical simulations
of stellar surface convection in late-type stars. This grid of simulations is
presented in more detail by \citet{trampedach:3Datmgrid}.
The first paper in this series
deals with the radiative part of the stellar surface problem and presents
{\Ttau}s derived from the simulations 
\citep[Paper\,I]{trampedach:T-tau}. The present paper is Paper\,II. Paper\,III
will address the consequences, of the results from the first two papers, for
stellar evolution (Trampedach et~al. in preparation).

In Sect.\ \ref{sect:MLT-3D} we outline some concepts of realistic 3D stellar
surface convection and contrast with the simple local mixing-length picture, which
nonetheless has some justification. In Sect.\ \ref{sect:sims} we introduce the
3D convection simulations and in Sect.\ \ref{sect:env}, the 1D envelope models.
Our calibration of the MLT mixing length, $\alpha$, by matching 1D models to
the deep part of our 3D simulations is described in
Sect.\ \ref{sect:env-match}. Results of the calibration are discussed in
Sect.\ \ref{sect:results}, including a note on gravity darkening.
The sensitivities of 1D structure models to the physics of the
outer boundary is explored in Sect.\ \ref{sect:StelStruc}, which also explains
the expansion of convective envelopes with increasing convective efficiency.
A more detailed discussion of our solar calibration is carried out in
Sect.\ \ref{sect:ObsCmp}, also including a perspective on 
semi-empirical calibrations for other stars from the literature.

We conclude in Sect.\ \ref{sect:concl} where we also provide instructions
for accessing our results on-line, and for using them in stellar structure
codes.

This paper is not a justification of the MLT, nor is it
aimed at describing the structure of the surface layers of stars.
Rather, we provide a way to use MLT and a non-constant $\alpha$ to 
determine the correct adiabat of the deep convection zone,
which in turn determines the depth of the convective envelope.
MLT in general, and our calibration of $\alpha$ in particular,
has limited relevance to stellar atmosphere calculations.

The present paper supersedes the work presented in
\citet{trampedach:score96}. The $\alpha$ calibration by
\citet[from here on LFS]{ludwig:alfa-cal}, against 2D simulations with grey
radiative transfer, was based on a method independent of ours. Their results
are similar to ours in many respects, as discussed in Sect.\ \ref{sect:results}.

\citet{asida:2DsimRG-alpha} performed a calibration against a 2D simulation of
a $36^\circ$ wedge of a spherical shell of a 1.2\,M$_\odot$ red giant, reaching
below the bottom of its convective envelope. It was truncated at the top, to
justify treating radiation in the diffusion approximation, and it was further
assumed that $\tau=2/3$ at the top boundary, and that the atmosphere is grey
so that $T_{\rm eff}$ is the local temperature there. This would seem
a limitation of the method, as convective efficiency is governed
by radiative cooling in the photosphere. A value of $\alpha=1.4$ is found from
a sign change in the initial flux transient, as function of $\alpha$ of the 1D
MLT models that provide the initial conditions. This 
seems to ignore the possibility of the interior adjusting differently
from the atmosphere and the former taking place on a much longer time-scale.

\citet{tanner:ZTtau}
evaluated the radius-$\alpha$ (or equivalently mass-$\alpha$) relation for MLT
stellar evolution models, restricted to have a prescribed set of atmospheric
parameters, $T_{\rm eff}$ and $\log g$. An age-$\alpha$ relation also results
from this and they also found the metallicity dependency of these relations.
They do not, however, employ results of their 3D simulations of convection,
treating radiation in the Eddington approximation.

\section{Mixing length versus 3D convection}
\label{sect:MLT-3D}

Comparisons between local mixing-length versions of convection and the results
of 3D simulations of the hydro-, thermo- and radiation-dynamics of convection
have been presented before
\citep{chan:3DtestingMLT,bob:conv-topology,freytag:2D-DA-WD+Astars+Sun,nordlund-stein:score96,abbett:3DvsMLT,Nordlund:3D-1D-conv,trampedach:Rome2009,trampeda:mixlength},
but we would like to point out some features that are pertinent to the present
calibration of the MLT mixing length.

The conventional interpretation of the mixing-length formulation of convection
is that of bubbles, eddies or convective elements that are warmer than their
surroundings, rising due to their buoyancy. The Schwarzschild criterion
for instability against convection
\be{conv-crit}
    \nabla_{\rm rad} > \nabla_{\rm ad}\ ,
\ee
is equivalent to the statement that convection will occur, when
gas which is warmer than its surroundings is buoyant. The logarithmic 
gradients, $\nabla={\rm d}\ln T/{\rm d}\ln p$, 
are labelled by the stratifications they apply to: ``rad'' for
radiative equilibrium and ``ad'' for adiabatic.

These bubbles of gas are then envisioned to travel for one
mixing length -- hence the name of the formulation -- before they dissolve
more or less abruptly
\citep{boehm:mlt}. This picture has conceptual problems at the edges of
convection zones or in small convective cores, where the distance to the edge
is only a fraction of a mixing length. Most often, the convective elements
are also ascribed an aspect ratio around unity, confounding the problem.

The mixing length is typically chosen to be
$\ell=\alpha H_\varrho$ or $\alpha H_p$, where $\alpha$
is the main free parameter (of order unity) of the formulation, and $H$ is
the density or pressure scale height for locally exponential stratifications.
It has also been suggested to use $\ell=\alpha z$, where $z$ is the distance
to the top of the convection zone
\citep{canuto-mazzitelli:conv,canuto-mazzitelli:conv-improv}. This choice
would solve the conceptual problems listed above, but it introduces physical
problems since there are strong reasons for real convection to have a
stratification similar to an MLT model with $\ell=\alpha H_\varrho$,
as mentioned below.

There is also a notion of these convective bubbles travelling in
a background of the average stratification.
The concept of a background liquid is rather obvious,
in this case, with isolated and distinguishable bubbles rising in it.

The 3D simulations of convection, on the other hand, display a very different
phenomenon \citep[see also][]{bob:conv-topology,nordlund-stein:score96,trampedach:Rome2009}.
The convection consists of continuous flows; the warm gas rising
almost adiabatically, in a background of narrower and faster downdrafts,
forced by sheer mass-conservation.
A fraction of the upflows is continuously overturning in order to conserve
mass on the back-drop of the steep and exponential density gradient.

Locally, the density, $\varrho$, as function if depth, $z$, can be approximated by
\be{exprho}
    \varrho(z) \propto e^{z/H_\varrho}\ .
\ee
When a vertical ``column''
of the upflow has travelled $\Delta z$, the column would therefore be
over-dense by a factor of $e^{\Delta z/H_\varrho}$, if the upflow were confined
horizontally. There is of course no such confinement, and the fraction,
$(e^{\Delta z/H_\varrho}-1) \rightarrow \Delta z/H_\varrho$ for
$\Delta z \rightarrow 0$, will overturn into the downdrafts.

The upflow will therefore be ``eroded'' by overturning, with an $e$-folding scale of
$H_\varrho$. The result of this concept is the same as that for
the mixing-length picture described above (with $\ell=\alpha H_\varrho$ and $\alpha=1$),
but without the same conceptual
problems. Since the flows are continuous, including the overturning of
upflows into the downdrafts, there are no discrete eddies of convection that
should be much smaller than the scale of changes in the atmosphere, as measured
by $H_p$. An $\alpha>1$ does therefore not pose a logical or physical
inconsistency, as argued by, e.g., \citet{canuto:2Dvs3D}. Many other
inconsistencies of MLT remain, however.

Renaming it the erosion- or dilution-length formulation, it could be a
first-order approximation to convection, as observed in the 3D simulations.
This is the reason that the MLT formulation has worked so well
despite its many short-comings: It is based on radial flows along a density
gradient, under the constraint of simple mass conservation.

%
%
The above argument neglects vertical velocity gradients. A positive outward
gradient accommodates more of the upflow and results in a smaller fraction
of the upflow overturning. Accounting for the velocity gradients present
in the simulations results in mixing lengths that are proportional to $H_p$
rather than $H_\varrho$, and discounts $\ell=\alpha z$, as shown by
\citet{trampeda:mixlength} for the same grid of simulations as used here. These actual
scale heights of mass mixing deviate drastically from this proportionality
in the surface layers, and are only loosely connected to the mixing length
of the MLT formulation being calibrated in the present paper.

The 3D simulations also display a nearly
laminar upflow, due to the density gradient smoothing out most of the
generated turbulence. The downflows are narrower and faster, and since
they work against the density gradient, they are also more turbulent.
The downdrafts are not compressed
adiabatically, since there is continuous entrainment of hot plasma from the
neighbouring upflows. The downdrafts therefore remain super adiabatic to
much greater depths 
than do the upflows which mainly become super adiabatic from radiative
loss of energy around the local photosphere, $\tau\simeq 1$. There
is also a lateral exchange of energy, extending the super-adiabatic peak in
the upflow to larger depth than would have been the case with a purely
vertical loss of radiative energy. This is also presented in Fig.~4 of
\citet{trampedach:Rome2009}.
The super-adiabatic peak produced by the
combination of these three phenomena is difficult to reproduce within the
%
%
MLT framework (See Sect.\ \ref{sect:solardCZ}).

Convective motions in the 3D simulations are prolific above the convection
zone, with the velocity decreasing with a scale height which is larger than
the pressure scale height. This introduces a new contestant in the atmosphere,
and radiative transfer will have to compete with adiabatic cooling 
for the equilibrium stratification, as discussed by
\citet{nordlund+stein:RadDyn1991}, \citet{asplund:low-Z-Li}, \citet{remo:3DRedGiantAbunds} and in Paper\,I.

The asymmetry between upflows and downdrafts have some profound effects:
In the photosphere, for example, the highly non-linear opacity, coupled with
the large temperature contrast, results in a visible surface which is very
undulated. Over the hot granules the $\tau\simeq 1$ surface is located at larger
geometrical height than in the cooler inter-granular lanes, and the observed
(disk-centre) temperature contrast is therefore much smaller than in a
horizontal cross-section \citep{georgobiani:p-mode-assym}. This higher opacity
in the granules causes a sub-photospheric heating of the granules (compared
with a 1D calculation) which is not balanced by a corresponding cooling below
the downflows, both because the opacity is convex in temperature and because
the downflows occupy a smaller fraction of the surface area.
This introduces a \emph{convective back-warming}
on the geometrical scale, which has no counterpart on the optical depth scale
\citep{bob:Tuebingen}.

In the convective layers, the density and velocity differences between the
upflows and the downdrafts give rise to a net kinetic energy-flux
\be{Fkin}
    F_{\rm kin} = \frac{1}{2}\varrho v^2 v_z\ ,
\ee
which amounts to 10--30\% of the total flux, depending on depth and stellar
parameters. The assumption of symmetry in the MLT formulation,
together with mass conservation, precludes such a kinetic energy-flux and
is probably the biggest cause for disagreement with the simulations in
the deeper, almost adiabatic part of the convection zone.

As convection approaches the adiabatic stratification, an actual
theory of convection is not needed in the bulk of convection zones
of more than a few pressure scale heights in depth. There we only need
to determine which adiabat the convection zone is following.
Since $\alpha$ is not
fixed by the MLT formulation, the answer has to come from ``outside''
calibrations, {\eg}, through the classical matching of the radius of a solar
model evolved
to the present age \citep{gough:MLT-calibr}, or as performed in the present
paper, against realistic 3D convection simulations.

We will refer in the following to $\alpha$ as a measure of the efficiency of
convection, but note here that the first-order term is already included in the
MLT formulation. That term is the ratio of convective advection of energy to
radiative diffusion of energy, known as the Pecl{\'e}t number which varies with
depth and is essentially what is solved for in the cubic MLT equation. The
primary MLT parameter, $\alpha$, acts as a modulation of that first order
convective efficiency provided by the Pecl{\'e}t number. Inherent in the MLT
formulations is therefore a change in efficiency across the HR diagram, and
within individual convective envelopes, with the least efficient convection to
be found at the surface of main-sequence stars so warm that the convective
envelope has almost disappeared. This, however, does not account for the fact
that in real convection, small changes in the asymmetry between up- and
downflows can greatly alter the transport properties of convection. Our
calibration of $\alpha$ displays an amplification of MLT's intrinsic decrease
of convective efficiency with increasing $T_{\rm eff}$ (see Sect.\
\ref{sect:results}).

Simulations in 2D are obviously attractive for their significantly lower
computational cost, compared with 3D simulations. We advice caution, however,
in interpreting both observations and phenomenological convection models in
terms of 2D simulations. The morphology of 2D and 3D convection is fundamentally
different, with 2D simulations being dominated by large-scale vortices which
also increases the size of granules
\citep{asplund:num-res,ludwig:2D-3DconvSims}. In 3D vortices become diluted by
the distance from their centres, and are consequently smaller and no longer the
dominant feature of the flows.

Other approaches to convection in stellar models include two-stream models
\citep{ulrich:convatm-2stream,aake:2comp,lesaffre:2streamConvModel},
plume models \citep{rieutord-zahn:plume-conv},
various closure models \citep{kuhfuss:t-depTurbConv,canuto:R-stress1,chan:turbconv4,kim:comp-mlt},
combinations of the latter two \citep{belkacem:PlumeClosureConv} and
semi-analytical models informed by simulations \citep{rempel:OvershootModel}.

\section{The simulations}
\label{sect:sims}
The fully compressible RHD simulations are described by \citet{trampedach:3Datmgrid}
and in Paper\,I, and the code
is described in more detail in \citet{aake:comp-phys}.
General properties of
solar convection, as deduced from the simulations, are discussed by 
\citet{stein:solar-granI}. Among the code features important for the present
analysis are radiative transfer with line-blanketing
\citep{aake:numsim1,stell-gran3},
and the transmitting top and bottom boundaries. The bottom is kept
at a uniform pressure (but not constant in time), to make a node in the radial
p modes and minimize
wave generation by the boundary conditions. The entropy of the inflowing plasma
is evolved towards a constant which is adjusted to result in the desired
effective temperature, and the velocities of the inflow are evolved towards 
a vertical flow that balances the mass-flux of the outflow.
The outflows are left unchanged in all respects.

Each of the simulations were performed on a $150\times 150\times 82$ grid,
equidistant and periodic in the horizontal directions, and optimized to resolve
the photospheric temperature gradient in the vertical direction. The top
boundary is located so that the maximum optical depth there is less than
$\log\tau\lwig -4.5$, and the bottom is located sufficiently deep for
convection to be largely adiabatic---about seven pressure scale heights below
the photosphere. The horizontal extent is chosen to cover of the order of
30 major granules. For the solar simulation this gives a simulation domain of
6\,Mm$\times 6$\,Mm$\times 3.5$\,Mm with 2.8\,Mm below the photosphere.

Convection in the simulations consists of a warm, coherent upflow, with
its entropy virtually unaltered from its value near the bottom of the 
convection zone.
This asymptotic value of the entropy in the deep convection zone is shown in
an \emph{atmospheric} HR-diagram (having $\log g$ as the vertical axis) in
Fig.~\ref{cwSmax} \citep[also Fig.~1 of][]{trampedach:3Datmgrid},
together with evolutionary tracks computed
with the MESA code \citep{paxton:MESA}.
This $S_{\max}$ is also the value of the entropy fed into the
simulations through the isentropic upflows at the bottom boundary.
\mfigur{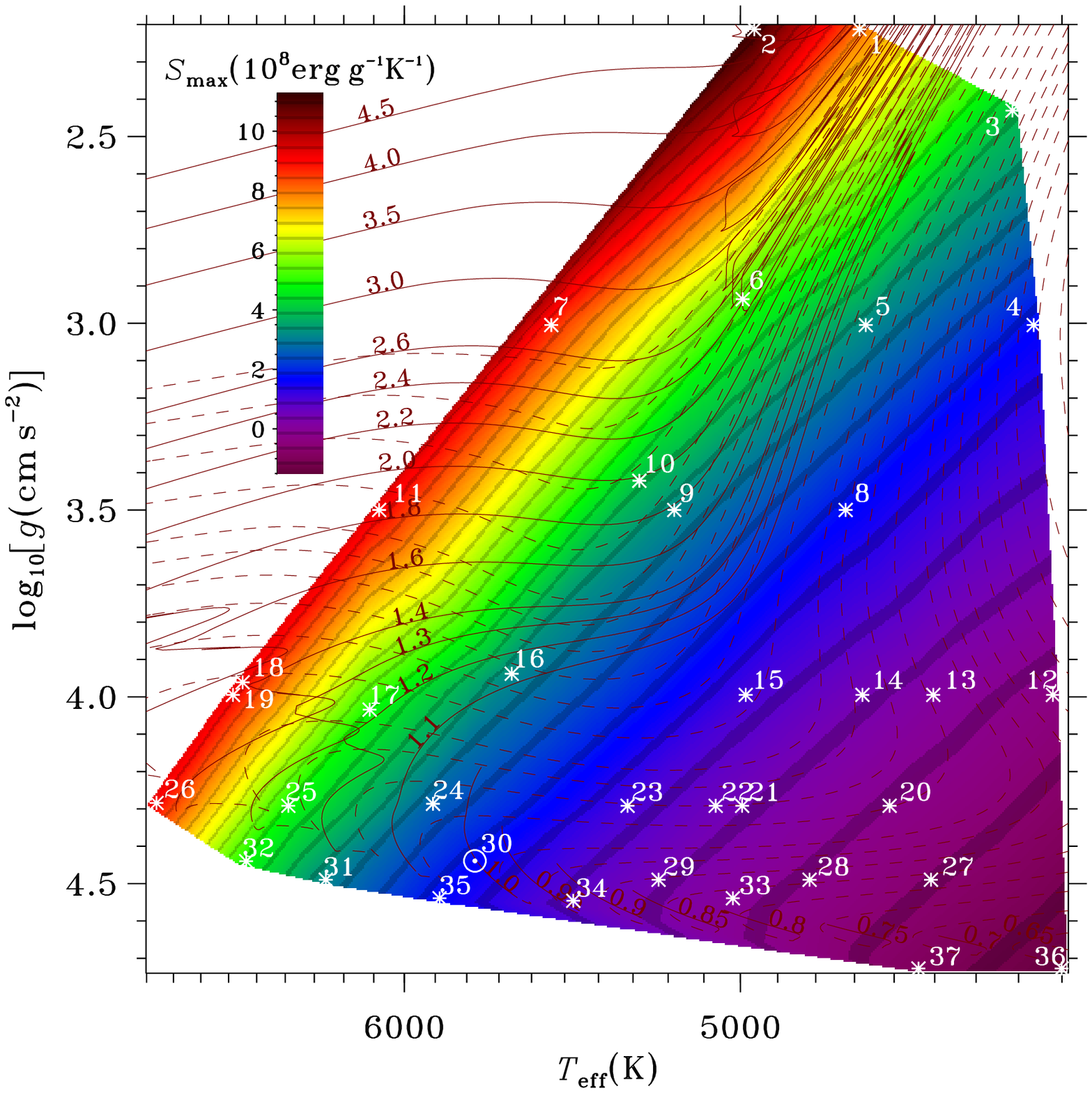}{8.7cm}
    {The asymptotic entropy (arbitrary zero-point),
     $S_{\rm max}/[10^8$\,erg\,g$^{-1}$K$^{-1}$], of the deep convection zone
     as function of stellar atmospheric parameters. This is also the entropy
     assigned to the inflows at the bottom of the simulation box.
     The $T_{\rm eff}$-scale
     is logarithmic. We have also plotted stellar evolution tracks produced
     with the MESA-code {\protect\citep{paxton:MESA}}, with masses as
     indicated, to put the simulations in context. The dashed part shows the
     pre-main-sequence contraction, and $\alpha$ and initial helium abundance,
     $Y_0$, were determined from a calibration to the present Sun.
     We interpolated the entropy linearly on Thiessen
     triangles \citep{renka:triangulation} between the simulations, and
     indicate the values 
     with colours as shown on the colour bar. The location of the simulations
     is shown with white asterisks, except for the solar simulation which is
     indicated with a $\odot$. For this figure we also added the
     simulation numbers from Table~\ref{starlist}.}
Because of the density gradient, mass conservation forces overturning
of the upflows, on distances of the order of the density scale height, as
mentioned in Sect.~\ref{sect:MLT-3D}. The
overturning plasma is entrained into narrow, fast and turbulent, entropy
deficient downdrafts, generated by the abrupt cooling in the photosphere. 
Since only a small part of the convection zone is simulated, open 
boundaries are necessary for obtaining realistic results. With closed
boundaries, the entropy deficiency and the turbulence would get recycled into the
upflows, artificially reducing the asymmetry between upflows and downdrafts.

Since spatial schemes of order higher than linear are unstable, artificial
diffusion is needed. It can be argued that a Laplacian term in velocity with
a constant viscosity would be the most physical choice, since it is the expected
form of molecular diffusion. On the other hand, the dissipation-scale of
atomic diffusion is many orders of magnitude removed from the smallest scales
resolved by the simulations, making that argument tenuous at best. A Laplacian
diffusion severely restricts the inertial range of hydrodynamic quantities,
thus greatly affecting the transport
properties of the plasma. Our simulations instead use a so-called
hyper-viscosity scheme, which is tailored to only affect regions where it is
needed and leave the rest untouched. This extends the inertial range without
introducing spurious artifacts, pushing the dissipation scale
towards the smallest scales resolved. In doing so, we loose contact with the
various dimensionless numbers of hydrodynamics, but we consider that a small
price to pay in return for a factor of about two in inertial range.
\citet{kyle:SSLinCSS} analysed a diffusion scheme conceptually similar to ours
(in that it minimizes diffusion everywhere except where it is needed for
stability) and compared with the simple Laplacian form. Similar to our case,
they found that at the same resolution, resolving power can be increased by
a factor of two when Laplacian diffusion is abandoned.

Another requirement for comparison with observations is a realistic 
treatment of the radiative
transfer in the atmosphere, and a corresponding quality of the atomic physics
behind the opacities and the EOS. Compared with the
simulations by \citet{aake:comp-phys}, 
we have therefore employed the so-called MHD EOS \citep{mhd1,mhd3}, updated
most of the continuous opacity sources and added a few new ones, as mentioned
in Paper\,I. The line opacity is
supplied by opacity distribution functions (ODF) by 
\citet{kur:line-data,kur:missolar}.

After relaxation to a quasi-stationary state, we calculated mean 
models for the envelope fitting ({\cf} Sect.\ \ref{sect:env-match}).
The temporal averaging was performed on a horizontally averaged column density
scale, instead of a direct spatial scale, to filter out the main effect of the
radial p modes excited in the simulations. We refer to this procedure as
\emph{(pseudo) Lagrangian averaging}, $\langle\dots\rangle_{\rm L}$.
A true Lagrangian average would be performed on the \emph{local} column mass
scale.

The simulations of the (irregular) grid are listed in Table~\ref{starlist} in
the same order as the corresponding tables in Paper\,I and
\citet{trampedach:3Datmgrid}, which is ascending in gravity and, for similar
gravities, ascending in effective temperature. 
The spectral type is only meant as a guide, and
none of our results relies on it. The simulations are run with constant gravity,
which in 1D is referred to as the plane-parallel approximation (as opposed to
spherical models), and are therefore independent of mass.
The stellar masses listed in Table~\ref{starlist} are therefore not based on the
simulations, but on the evolution tracks shown in Fig.~\ref{cwSmax}. They are,
however, used for the $\alpha$ calibration in Sect.\ \ref{sect:env-match}, but
the results are insensitive to the choice of mass with $\upartial\alpha/\upartial
M \simeq 7\stimes 10^{-4}$, with mass in units of M$_\odot$.
The chemical
composition is a modified \citet{AG89} metal mixture, with helium and iron
adjusted to correspond to the \citet{GS98} mix. This results in $X=73.70$\,\%
hydrogen by mass and $Y=24.50$\,\% helium by mass, the latter in accordance
with helioseismology \citep{jcd+perez:santa-barbara,dziembowski:SunSeismHe,vorontsov:SunSeismHe,basu-antia:Y,richard:sunHe}. This also constitutes (1.26, 1.52 and
1.55 times) higher
abundances of C, N and O than found in the solar abundance analysis by
\citet{AGSS2009} based on a 3D solar simulation similar to ours. These
elements, however, provide little opacity under the physical conditions of our
simulations and we expect the abundance differences to have minor effects on
our results, as also mentioned in Sect.\ \ref{sect:solardCZ} (the source of the
so-called solar abundance problem \citep{bahcall:seism-AGS05} lies in the
importance of the O opacity at the bottom of the solar convection zone). The
detailed abundances are listed and further discussed by
\citet{trampedach:3Datmgrid}.

\aatab{starlist}{Fundamental parameters for the 37 simulations,
   and calibrated $\alpha$'s and convection zone depths, $d_{\rm cz}$, in units of stellar radius. The standard deviations describe the fluctuations in time.}{rcccccc}
{ \oh{sim} & \oh{type} & \oh{$T_{\rm eff}$/[K]}& \oh{$\log g$}& \oh{$M/$M$_\odot$}& \oh{$\alpha$}& \oh{$d_{\rm cz}$} \\}{
  1 &     K3 & $4\,681\pm 19$ & 2.200 & 3.694 & $1.673\pm 0.024$ & 0.49555 \\
  2 &     K2 & $4\,962\pm 21$ & 2.200 & 4.805 & $1.559\pm 0.025$ & 0.28706 \\
  3 &     K5 & $4\,301\pm 17$ & 2.420 & 0.400 & $1.750\pm 0.021$ & 0.97807 \\
  4 &     K6 & $4\,250\pm 11$ & 3.000 & 0.189 & $1.807\pm 0.020$ & 0.99980 \\
  5 &     K3 & $4\,665\pm 16$ & 3.000 & 0.852 & $1.750\pm 0.020$ & 0.91381 \\
  6 &     K1 & $4\,994\pm 15$ & 2.930 & 2.440 & $1.705\pm 0.021$ & 0.51661 \\
  7 &     G8 & $5\,552\pm 17$ & 3.000 & 2.756 & $1.627\pm 0.021$ & 0.22159 \\
  8 &     K3 & $4\,718\pm 15$ & 3.500 & 0.721 & $1.746\pm 0.023$ & 0.81835 \\
  9 &     K0 & $5\,187\pm 17$ & 3.500 & 1.786 & $1.760\pm 0.028$ & 0.51630 \\
 10 &     K0 & $5\,288\pm 20$ & 3.421 & 1.923 & $1.723\pm 0.029$ & 0.43090 \\
 11 &     F9 & $6\,105\pm 25$ & 3.500 & 1.875 & $1.638\pm 0.027$ & 0.11782 \\
 12 &     K6 & $4\,205\pm ~8$ & 4.000 & 0.601 & $1.994\pm 0.032$ & 0.76238 \\
 13 &     K4 & $4\,494\pm ~9$ & 4.000 & 0.684 & $1.838\pm 0.018$ & 0.65469 \\
 14 &     K3 & $4\,674\pm ~8$ & 4.000 & 0.738 & $1.779\pm 0.024$ & 0.60464 \\
 15 &     K2 & $4\,986\pm 13$ & 4.000 & 0.836 & $1.755\pm 0.022$ & 0.52884 \\
 16 &     G6 & $5\,674\pm 16$ & 3.943 & 1.130 & $1.756\pm 0.034$ & 0.34300 \\
 17 &     F9 & $6\,137\pm 14$ & 4.040 & 1.222 & $1.697\pm 0.025$ & 0.19127 \\
 18 &     F4 & $6\,582\pm 26$ & 3.966 & 1.567 & $1.655\pm 0.020$ & 0.06244 \\
 19 &     F4 & $6\,617\pm 33$ & 4.000 & 1.552 & $1.659\pm 0.020$ & 0.05981 \\
 20 &     K4 & $4\,604\pm ~8$ & 4.300 & 0.568 & $1.819\pm 0.023$ & 0.50541 \\
 21 &     K1 & $4\,996\pm 17$ & 4.300 & 0.694 & $1.734\pm 0.022$ & 0.44249 \\
 22 &     K1 & $5\,069\pm 11$ & 4.300 & 0.719 & $1.739\pm 0.027$ & 0.43124 \\
 23 &     K0 & $5\,323\pm 16$ & 4.300 & 0.810 & $1.742\pm 0.024$ & 0.39198 \\
 24 &     G1 & $5\,926\pm 18$ & 4.295 & 1.056 & $1.752\pm 0.029$ & 0.26388 \\
 25 &     F5 & $6\,418\pm 26$ & 4.300 & 1.261 & $1.715\pm 0.030$ & 0.13408 \\
 26 &     F2 & $6\,901\pm 29$ & 4.292 & 1.433 & $1.684\pm 0.038$ & 0.03882 \\
 27 &     K4 & $4\,500\pm ~4$ & 4.500 & 0.565 & $1.874\pm 0.016$ & 0.41811 \\
 28 &     K3 & $4\,813\pm ~8$ & 4.500 & 0.664 & $1.752\pm 0.022$ & 0.38241 \\
 29 &     K0 & $5\,232\pm 12$ & 4.500 & 0.812 & $1.741\pm 0.020$ & 0.33902 \\
 30 &     G5 & $5\,774\pm 17$ & 4.438 & 1.002 & $1.764\pm 0.030$ & 0.27910 \\
 31 &     F7 & $6\,287\pm 15$ & 4.500 & 1.246 & $1.761\pm 0.036$ & 0.16503 \\
 32 &     F4 & $6\,569\pm 17$ & 4.450 & 1.329 & $1.708\pm 0.021$ & 0.10530 \\
 33 &     K1 & $5\,021\pm 11$ & 4.550 & 0.772 & $1.760\pm 0.024$ & 0.34027 \\
 34 &     G9 & $5\,485\pm 14$ & 4.557 & 0.949 & $1.760\pm 0.024$ & 0.29097 \\
 35 &     G1 & $5\,905\pm 15$ & 4.550 & 1.114 & $1.770\pm 0.027$ & 0.23625 \\
 36 &     K6 & $4\,185\pm ~3$ & 4.740 & 0.649 & $2.050\pm 0.027$ & 0.33324 \\
 37 &     K4 & $4\,531\pm 10$ & 4.740 & 0.742 & $1.918\pm 0.040$ & 0.29067 \\
}T

\mfigur{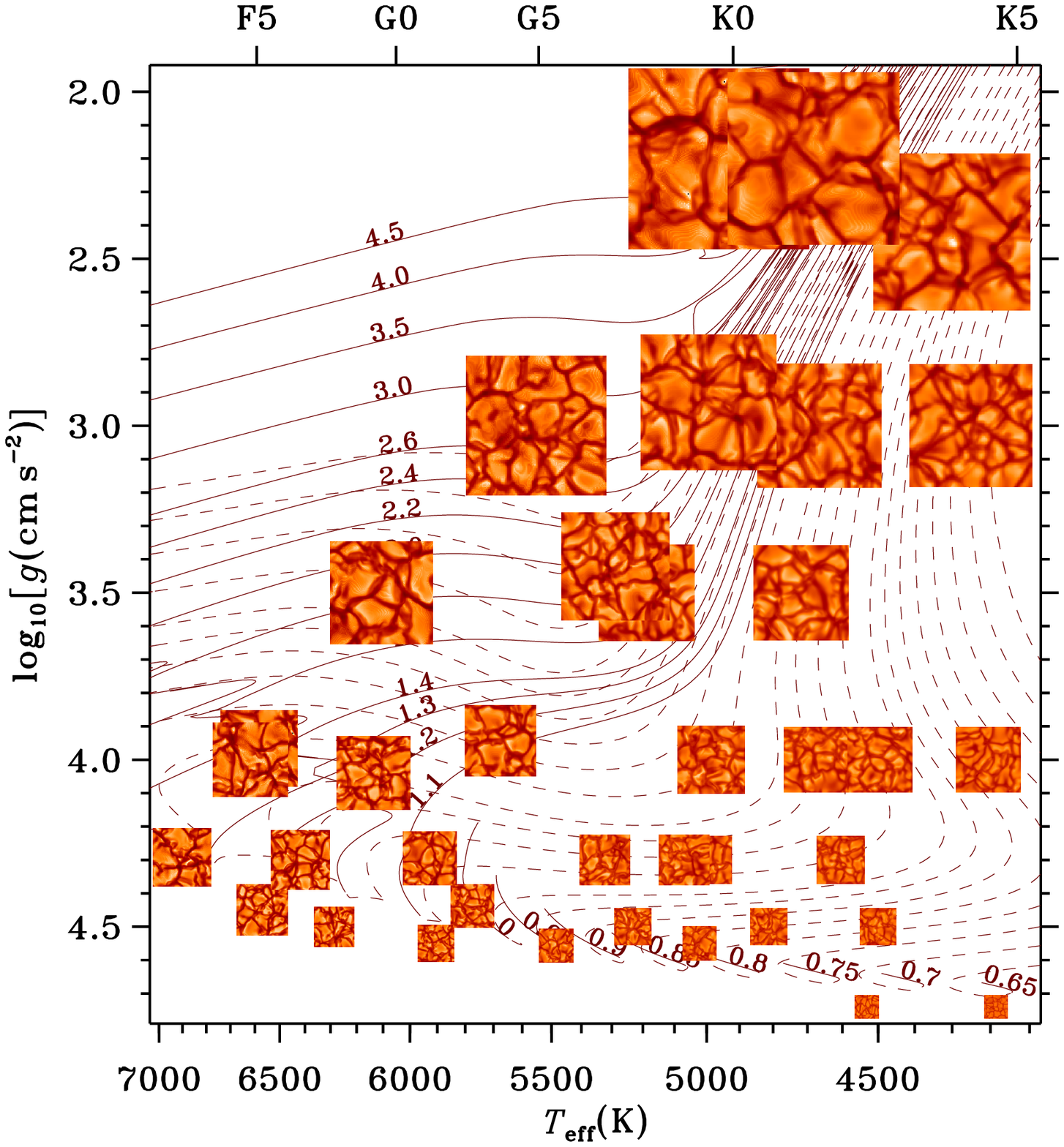}{9.0cm}
    {The position of the simulations in the atmospheric HR-diagram (with a
     logarithmic $T_{\rm eff}$-axis).
     The simulations are represented by random snapshots of the disk-centre
     white light intensity.
     The size of these snapshots is proportional to the logarithmic width
     of the simulation domains.
     The evolutionary tracks are the same as shown in Fig.~\ref{cwSmax}.}

Fig.~\ref{cHR_g} presents our simulation grid in the atmospheric
HR-diagram, showing granulation snapshots of the simulations.
The randomly selected snapshots are scaled by the logarithmic horizontal
extent of each of the simulations---an extent which in turn is chosen to
encompass about 30 major granules. The 
evolution tracks from Fig.~\ref{cwSmax} are added for reference.

\section{The envelope models}
\label{sect:env}

The simulations were fitted to 1D, spherically symmetric envelope models
\citep{jcd-srf:rad-osc} computed with a code closely related to the ASTEC
stellar evolution code \citep{jcd:ASTEC}. The envelopes extend
from a relative radius of $r/R=0.05$, and
out to an optical depth of $\tau=10^{-4}$.

We used the same MHD EOS, and in the atmospheric part of
the envelopes we used the same opacities, as in the convection simulations (see
Paper\,I, Sect.\ 3.1).
These atmospheric opacities are smoothly joined with the updated
OP opacities \citep{OP05} between temperatures of $\log T= [3.95; 4.25]$ (where the
differences are small and the models have nearly adiabatic convection; see also
Fig.~2 of Paper\,I).
These interior opacities were computed for the
exact same composition as the atomic physics for the simulations as listed in
Table~1 of \citet{trampedach:3Datmgrid}. This combined set of atmospheric and
interior opacities are now included in the {\tt OPINT} opacity interpolation
package by \citet{houdek:gh_int.v6}, both of which can be downloaded from
\url{http://phys.au.dk/~hg62/OPINT}.

Convection is treated using the standard MLT as described in
\citet{boehm:mlt}, using the standard mixing length
\be{alpha}
    \ell=\alpha H_p\ ,
\ee
as supported by analysis of the same simulations by \citet{trampeda:mixlength}.
We use form factors $\Phi=2$ and $\eta=\sqrt{2}/9$ according to the notation of
\citet{gough:MLT-PulsStars}, with the values chosen to reproduce the
original formulation by \citet{boehm:mlt}\footnote{She brought up several values for the various form factors, but it is made clear which values she adopted in the end.}.
[See \citet[App.~A]{ludwig:alfa-cal} for more details about the form factors. We
use $f_1=\sfrac{1}{8}$, $f_2=\sfrac{1}{2}$, $f_3=24$ and $f_4=0$ in their
notation].

The photospheric transition from optically thick to optically thin 
is treated by means of
{\Ttau}s derived from the simulations in Paper\,I. There we calculated temporal
and $\tau$ (Rosseland optical depth) averaged temperatures, and
reduced them to radiative equilibrium, $T_{\rm rad}$, and computed generalised Hopf functions,
\be{eq:hopf}
    q(\tau) = \frac{4}{3}\left(\frac{T_{\rm rad}(\tau)}{T_{\rm eff}}\right)^4-\tau\ ,
\ee
as detailed in Paper\,I. This new formulation of the outer boundary ensures
consistency between the {\Ttau} and the 1D model it is implemented in. It also
applies throughout the stellar model, abandoning the artificial
distinction between the atmosphere and the interior of a structure model, and
renders moot the issue of where to switch between the two.
We interpolate linearly in $q$ between the simulations using
a Thiessen-triangulation \citep{renka:triangulation} of the irregular grid in
$\log_{10}T_{\rm eff}$ and $\log_{10}g$.
The point that we use individual {\Ttau}s instead of scaled
solar {\Ttau}s is crucial for the present calibration, as
discussed in Sect.\ \ref{sect:StelStruc}.

All time-dependent and composition-altering processes, {\eg}, nuclear reactions,
diffusion and settling of helium and metals, are left out of envelope models.
This renders the envelopes functions of the atmospheric parameters,
$T_{\rm eff}$ and $g_{\rm surf}$ (and composition) only,
but it also rules out any abundance gradients.
During stellar evolution, on the other hand, the net effect of radiative
acceleration and gravitational settling, is for
helium and metals to slowly drop out of the convection zone \citep{turcotte:Diff+RadLev}
building up an
abundance gradient, just below the convection zone, which is smoothed out by
chemical diffusion.
As these processes are much slower than convection, the resulting abundance
gradients are confined to below the convection zone, leaving our $\alpha$
calibration unaffected. The resulting depths of convection zones, on the other
hand, \emph{will} be affected by such abundance gradients.
Radiative levitation in the atmosphere \citep{hui-bon-hua:grad-staratm} would
segregate atoms and ions according to opacity, if it was not for the
convective overshoot we see in our simulations, sustaining appreciable velocity
fields and ensuring complete mixing, {\em at least} out to $\log\tau =-4.5$.

The pressure in the simulations is not purely thermodynamic;
turbulent pressure also contributes to the hydrostatic equilibrium. We
therefore include a turbulent pressure in the envelopes, based on
the MLT convective velocities \citep[see also][]{stellingwerf:MLT-Pturb}
\be{Pturb1D}
    p_{\rm turb,1D} = \beta v^2_{\rm conv}\varrho\ ,
\ee
where $\beta$ is a constant, adjusted as part of the calibration procedure,
described in Sect.\ \ref{sect:env-match}. 
For the calibration, we only need $p_{\rm turb,1D}$ from the matching point
and into the interior, and we suppress it smoothly in the layers above.
This has two reasons:
The practical one is that most stellar structure calculations do not include
such a turbulent pressure, and a calibration of $\alpha$, including
$p_{\rm turb,1D}$ in the whole convection zone, would not apply in these cases.
The second, and more important reason, is that $v_{\rm conv}$ in MLT models
displays a very sudden drop to zero at the top of the convection zone. In the
standard solar model S of \citet{GONG-Sci:sol-mod}, the drop from a peak of
4\,km\,s$^{-1}$ occurs over just
72\,km, corresponding to 0.42 pressure scale heights, $H_p$. Such a large
velocity gradient will give rise to a devastating
pressure gradient, causing nonphysical and sizable inversions in both density
and gas pressure. This is shown in Fig.~\ref{ptplot} for our simulation with
the most vigorous convection, the 6\,900\,K dwarf, No.\ 26 in
Table~\ref{starlist}. From that figure we see that not suppressing $p_{\rm turb}$
\mfigur{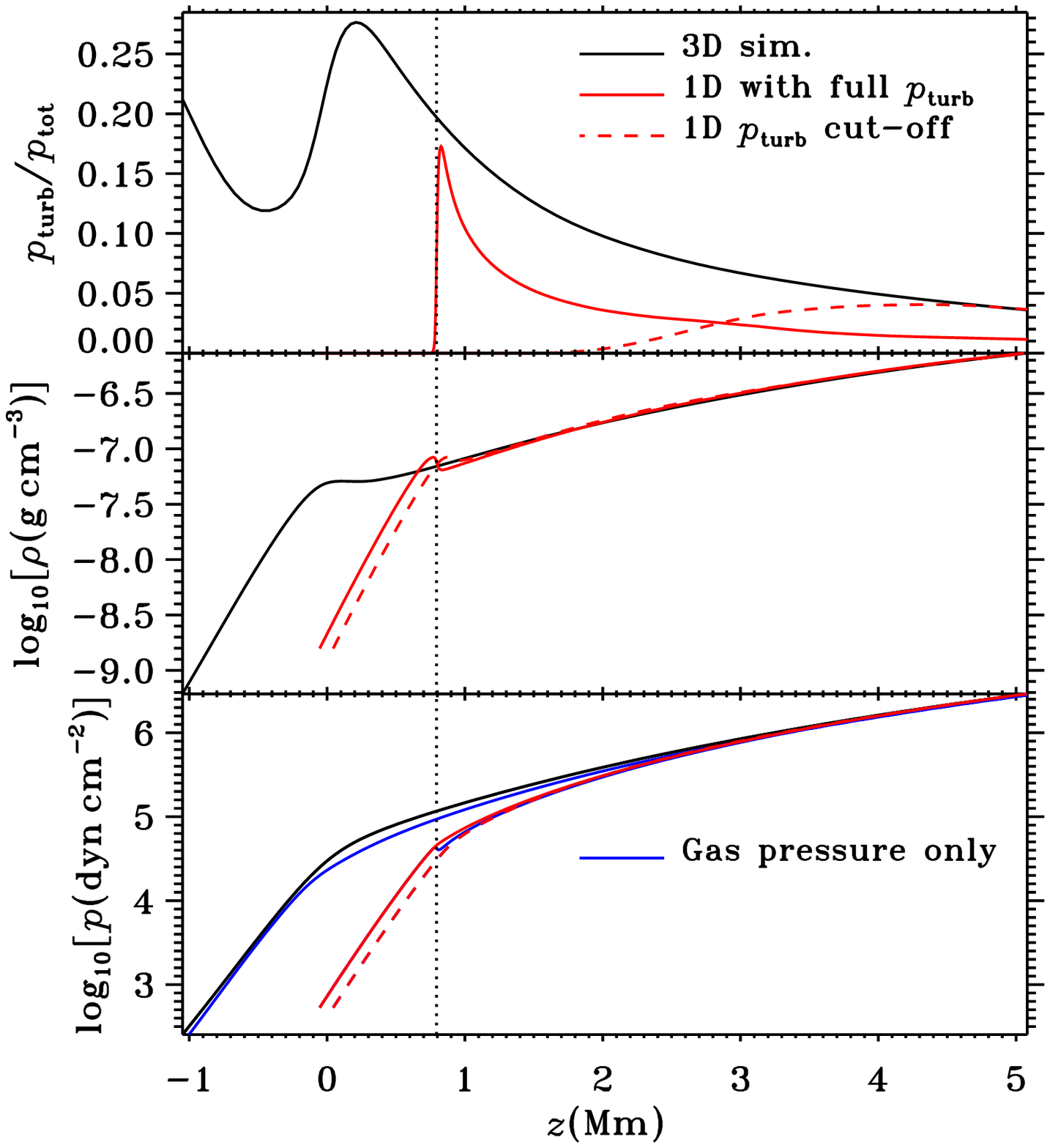}{9.5cm}
    {Comparing the structure of the $T_{\rm eff}=6\,901$\,K, $\log g=4.292$
     3D convective atmosphere in solid black, and some 1D counterparts in red.
     The three panels show turbulent to total pressure ratio, logarithmic
     density and logarithmic pressures as function of depth. The red dashed
     lines show the 1D envelope model calibrated to the 3D simulation. The
     turbulent pressure is suppressed towards the surface, as is evident from
     the top panel. The solid red line shows a similar 1D model, but without
     this suppression of $p_{\rm turb}$, and with $\beta=0.260$, as opposed to
     the calibrated 0.804. In the bottom panel the gas pressure is shown in
     blue.}
at the surface causes a gas pressure inversion, not seen in the simulations, as
well as a much increased density inversion. To include this full 1D turbulent
pressure, we had to limit the form-factor of equation~(\ref{Pturb1D}) to
$\beta=0.260$, as opposed to $\beta=0.804$ from the calibration. This is seen
in the top panel as a lack of convergence to the 3D result with depth. If a
model with the more realistic value of $\beta$ could converge, the turbulent
pressure would be the dominant contributor, at 53.4\% of the total pressure.
The local pressure scale height at the turbulent pressure peak is about
0.8\,Mm, and the drop from that peak to zero occurs in under 15\% of that. To
enable integration of hydrostatic equilibrium, it is therefore necessary to
introduce some cut-off for $p_{\rm turb,1D}$. We have constructed this to be
smooth and gradual (small gradient) and deep, just above the matching point,
minimizing its effect on the models.

In the 3D simulations, on the other hand, turbulent pressure peaks about half a
pressure scale height below the top of the convection zone, drops off smoothly
both above (from convective overshooting) and below and is non-zero at all depths.
The maximum (with depth) of the turbulent to total pressure ratio
is shown in Fig.~3 of \citet{trampedach:3Datmgrid}, and that quantity is a
good measure of how vigorous the convection is. For the Sun, that ratio is
13.75\%.

We recommend
not including a 1D-turbulent pressure which is confined to the convection zone,
since it will exhibit unphysically large and damaging gradients at the top
of the convective envelope (see Fig.~\ref{ptplot}).
\citet{baker:PulsConvRRLyr} noted that the turbulent pressure, if included
consistently (in all terms), increases the order of the equations for the
structure of the envelope model from three to four. They further found that
without overshoot the transition to the non-convective case at the boundaries
of the convection zone gives rise to singularities which have proved very
difficult to treat numerically.
Most non-local and/or time-dependent, MLT-style descriptions of convection
also accommodate overshooting
\citep{spiegel:non-localMLT1,shaviv:SunOvershootNL-MLT,aake:on_conv,ulrich:nonlocMLT1,gough:MLT-PulsStars,eggleton:consistentConv,grossman:nonloc-conv3,grigahcene:perturbMLT}, and some of these have the correctly negative convective
flux in the overshoot region. Most of these still have a strong and direct
connection between velocities and fluxes, meaning that the velocities still 
fall off to sharply, producing unphysically large gradients in turbulent
pressure, or alternatively, producing too extended a flux transition from
convection to radiation
\citep[see also][ for further analysis]{renzini:OvershootCritique}.
The reason for this is the assumed constancy of the fractional area occupied
by each direction of convective flows, $f_{\rm up}$ and $f_{\rm dn}$, and the
often assumed symmetry of these: $f_{\rm up}=f_{\rm dn}=1/2$. Some of these models
have positive convective fluxes in the overshoot region because they mainly
consist of a smoothing of quantities over some length scale.
In our simulations the negative convective overshoot flux is due to a
reversal in temperature change in response to a radial displacement, with the
upflows getting cooler than the downflows. 
Our simulations do not address the issue of overshooting at the bottom of
convective envelopes, but one could argue that a successful model for
overshooting should also be able to reproduce, at least qualitatively, our
results for overshoot into the atmosphere. Overshooting from convective cores
and the bottom of convective envelopes is crucial for various stages of stellar
evolution and an improved understanding is much needed.

Well below the super-adiabatic top of the convection zone, $p_{\rm turb,1D}$
does
match the turbulent pressure of the simulations rather well, giving
an almost differentiable match. This is part of our evidence that envelope models
including $p_{\rm turb,1D}$, with $\beta$ and $\alpha$ fitted as described in
Sect. \ref{sect:env-match}, give a realistic extension of the simulations towards
the centre of the star. This fact was exploited in an investigation
of convective effects on the frequencies of solar oscillations
\citep{rosenthal:conv-osc} by analysing eigenmodes in a model combining the
simulation and a matched envelope model. This same procedure 
was also employed in a
calculation of p-mode excitation \citep{bob:IAU-GA2003Sydney,bob:IAU-Symp239}
for a range of stars.
%
We can now proceed with the matching, with confidence.

\section{Matching to envelope models}
\label{sect:env-match}
In order to derive $\alpha$ values from the simulations we matched
1D envelope models to horizontal and temporal averages of the 3D simulations
at a common pressure point deep in the simulation.
The matching is performed by adjusting $\beta$, equation~(\ref{Pturb1D}), until the 1D-turbulent
pressure agrees with that of the simulation, and $\alpha$ until the temperatures agree, while keeping the
mass and luminosity of the envelope constant. The masses, as listed in Table~\ref{starlist},
are chosen based on the atmospheric parameters of evolution tracks constructed
with the MESA-code \citep{paxton:MESA}, calibrated to the present Sun as per
usual \citep{gough:MLT-calibr}. The luminosity follows from this mass and the
atmospheric parameters of the simulations. These masses are obviously
inconsistent with our calibration, but our calibration is also insensitive to
this choice of mass, with $\upartial\alpha/\upartial M\sim 0.001$.

This method demands a high degree of consistency between the simulations and
the envelope models at the matching point, which is the reason for using the
exact same EOS (and chemical composition) in both cases, and for including
a turbulent pressure in the deep part of the envelope models. 
The depth of the matching point is chosen for each simulation as a trade-off
between minimizing boundary effects (which increase with depth), and minimizing
fluctuations in thermodynamic quantities (which decrease with depth).
The latter is to ensure that the mean $\varrho$, $T$ and $p_{\rm gas}$
are related by the EOS, {\ie}, that direct 3D-effects are negligible, as is
of course always the case in 1D models. In all our cases, the matching point
is located deeper than $\log\tau=4$, and at pressures at least 100 times larger
than in the photosphere.

In order to filter out non-convective effects from this calibration of $\alpha$,
we also demand consistency between 1D and 3D in the treatment of radiative
transfer in the atmosphere. We accomplish this by using the Rosseland opacities
and the {\Ttau}s from the 3D simulations, in the atmospheres of the 1D envelope
models (see Paper\,I for details).

We also note that the common pressure point between the averaged 3D simulation
and 1D model, being located deep in the simulation, means that the two models
generally will disagree on the location of the surface, $R$, of the star (where
$\langle T\rangle = T_{\rm eff}$). This is due to the convective expansion of 3D
atmospheres compared with 1D, as discussed by \citet{trampedach:3Datmgrid}. The 1D envelopes
therefore have slightly smaller radii (by up to a per cent of the stars radius for the warmest giant and by as little as $10^{-5}$ for the coolest dwarf), and hence slightly larger $T_{\rm eff}$
and $\log g$, when the 3D simulation and 1D envelope are constrained to have
identical mass and luminosity.

The main advantage of our method for calibrating the MLT $\alpha$ is the resulting
combined models of averaged 3D simulations outside the matching point and
the calibrated 1D envelope model interior to the matching point. Our method
ensures these combined models are continuous, and they can therefore be used
for computing various asteroseismic quantities, which can then be interpolated
in the grid of simulations. Such combined models have been used by
\citet{rosenthal:conv-osc} to estimate the helioseismic surface effect, i.e.,
systematically overestimated model frequencies due to shortcomings at the
surface of 1D solar models. They have also been employed in a
calculation of p-mode excitation for a range of stars
\citep{bob:IAU-GA2003Sydney,samadi:ExciComp,bob:IAU-Symp239}, based on the
formulation by \citet{aake:osc-conv,bob:mode-exci}. Such calculations are
also planned for our grid of simulations and $\alpha$ calibrated 1D models
presented here.
The method used by LFS, of matching to the
entropy of the adiabat, does not ensure such continuous matching of 1D interior
and averaged 3D atmospheres and complicates their use for asteroseismic
applications. Their method does, however, ensure the correct location of the
bottom of the convective envelope as determined by the Schwarzschild criterion. 
The two methods will converge with deeper matching point of our method, as
the stratification approaches the adiabat exponentially.

\section{Results}
\label{sect:results}

In Fig.~\ref{cwalfa} we show the results of this calibration of the MLT $\alpha$
as function of effective temperature and gravity. The corresponding
atmospheric entropy jump is shown in Fig.~4 of \citet{trampedach:3Datmgrid}.
We also list these calibrated $\alpha$ values in Table~\ref{starlist} together
with their standard deviations, from performing the envelope matching to
individual time-steps of the horizontally averaged simulations.
This scatter in $\alpha$ ranges between 0.015 and 0.040, with an average of 0.026.
The standard deviations listed for
$T_{\rm eff}$ is likewise from the fluctuations in time. In Table~\ref{starlist}
we also list
the resulting depths of the convective envelopes, $d_{\rm cz}$, relative to the
stellar radii.
\mfigur{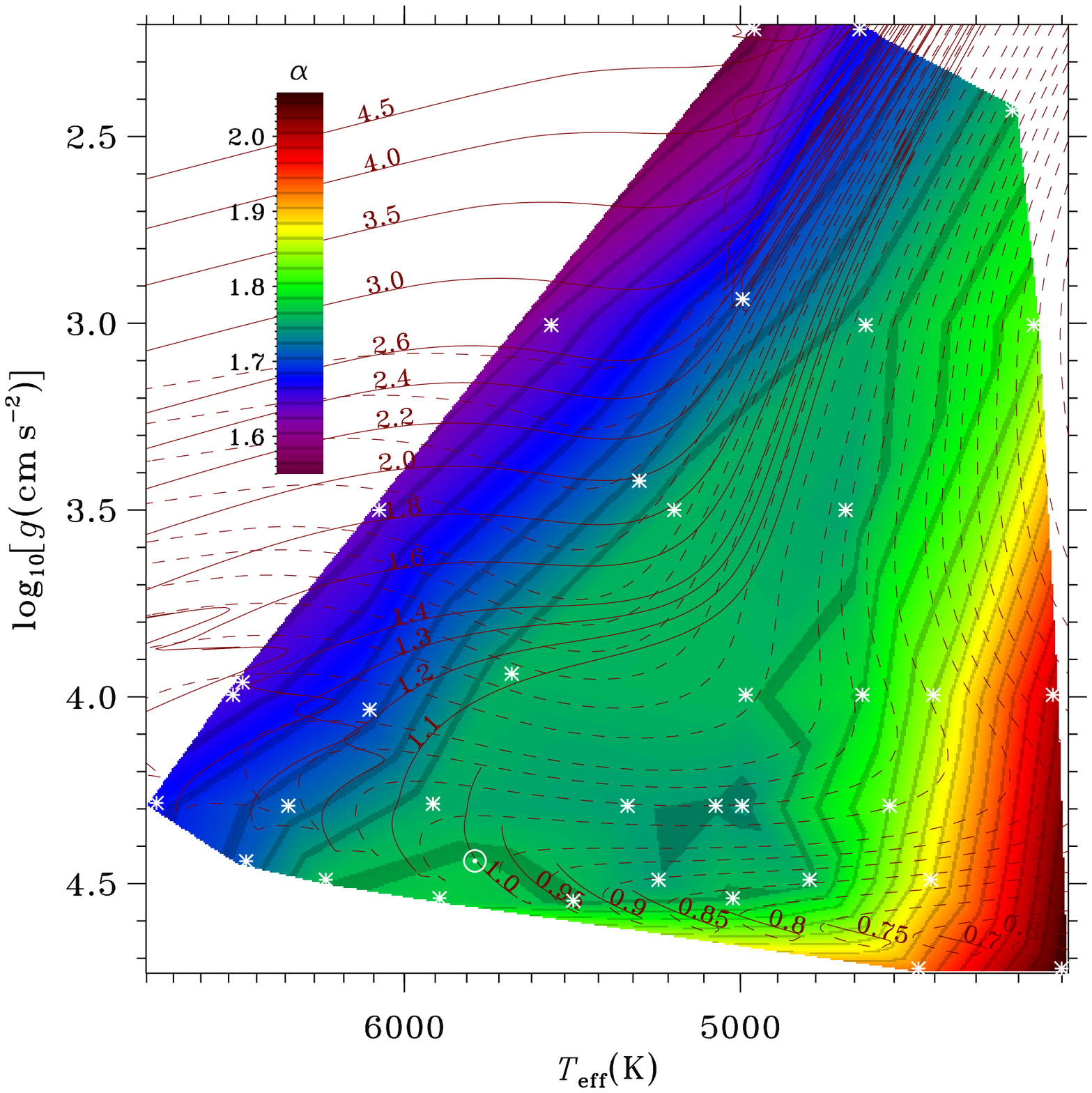}{8.7cm}
    {As Fig.~\ref{cwSmax}, but showing the behaviour of $\alpha$ with $T_{\rm eff}$
     and $g_{\rm surf}$. The over-plotted evolutionary tracks cover the
     mass-range 0.65--4.5\,M$_\odot$, as indicated.
     The solar simulation is indicated with a $\odot$ and the locations of the
	 other simulations, as listed in Table~\ref{starlist}, are shown with
     asterisks.}%

In Fig.~\ref{hgl_2dalfa} we show our calibrated $\alpha$ values
as function of $T_{\rm eff}$, with
error-bars corresponding to the RMS scatter in $T_{\rm eff}$ and $\alpha$.
The local $\log T_{\rm eff}$-gradient of $\alpha(T_{\rm eff}, g)$
is indicated with line-segments,

Our $\alpha$ calibration results in low values along the warm edge of our grid,
which is approaching the end of convective envelopes from the cool side of the
HR diagram. The
depth of convective envelopes decreases as this edge of our grid is approached, as
does the convective efficiency, quantified by $\alpha$. The highest
$\alpha$ values are found in the coolest dwarfs, spanning the range from
$\alpha = 1.68$ to 2.05 on the main sequence. A 0.7\,M$_\odot$-star would
experience the largest change in $\alpha$ over its life-time (about 0.15),
whereas 0.8--1.1\,M$_\odot$ stars find themselves on
a triangular plateau of $\alpha\sim 1.76$, spanning a temperature range of $T_{\rm eff}\sim 4\,800$--6\,000\,K 
along the main sequence and going up to $\log g\simeq 3.0$.
This feature was also noted by LFS. This plateau encompasses the evolution
so far of the Sun, as well as the nearest and best constrained
binary, $\alpha$\,Cen\,A and B (the C component is too cool to be covered
by our grid of simulations), which means that the
evolution of these three stars is well described by a single value of
$\alpha=1.76\pm 0.01\pm 0.03$ ($\pm$range, $\pm$uncertainty of calibration).

Stars ascending the giant branch essentially follow iso-$\alpha$ curves, so
here $\alpha$ is a function of mass and approximately constant with age.
The values here
range from 1.56 for a 4.3\,M$_\odot$-star to 1.75 for 0.84\,M$_\odot$.
Stars more massive than about 1.25\,M$_\odot$ experience a significant increase
in $\alpha$ during their evolution, but most of that change occurs in the
Hertzsprung gap of fast evolution. Effectively they have a small $\alpha$ on
the main sequence and a larger $\alpha$ on the giant branch. Bear in mind,
however, that our grid of simulations only covers up to 1.4\,M$_\odot$ on the
main sequence.

\mfigur{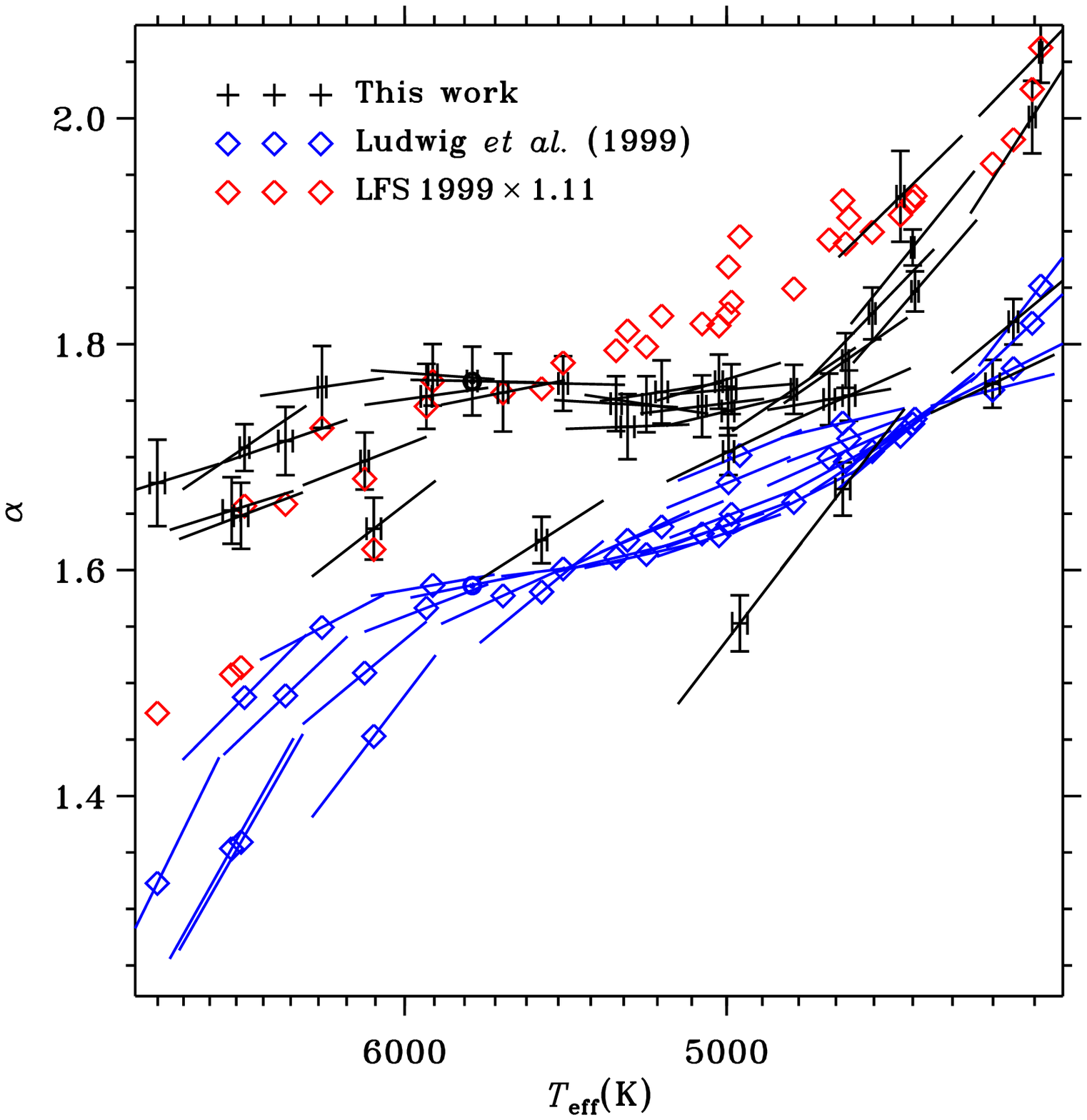}{9.3cm}
   {A plot of the values of $\alpha$ found from our matching procedure (with
    gradient-lines, showing the local slope in $\log T_{\rm eff}$, and
    error-bars in both $T_{\rm eff}$ and $\alpha$), compared with the calibration against 2D simulations,
    performed by {\protect\citet{ludwig:alfa-cal}} (lower, blue diamonds with
    gradient lines).
    We have also multiplied their result by 1.11 to agree
    with our result for the Sun (upper, red diamonds, no gradient lines).}%

\subsection{Comparison with \citet{ludwig:alfa-cal}}
\label{sect:LFScmp}

The calibration of $\alpha$ against 2D RHD simulations performed by LFS
employed a method completely independent of ours.
In Fig.~\ref{hgl_2dalfa} we have displayed their fit to their results,
as applied to the atmospheric parameters of our simulations
(lower set of blue $\diamond$'s, with line-segments indicating their local
$\log T_{\rm eff}$-derivatives).
They suggest a scaling of their results by 1.1--1.2 to translate from 2D to 3D,
and we note the profound morphological differences between
2D and 3D atmospheric convection, as discussed by \citet{asplund:num-res}.
LFS also recommend that in order to match 1D solar structure models to the
present age Sun, modellers should use the LFS $\alpha$ calibration
differentially, scaling it by their $\alpha_\odot$.
We fully adopt that recommendation, with the caveat that the {\Ttau}s and
atmospheric opacities of the 3D simulations should be used alongside our
$\alpha$ calibration. We expect this procedure to result in a minimal need
for scaling, but would be most interested in hearing of contrary experiences.

Our results do indeed agree with LFS's in the solar vicinity, after
a scaling by 1.11, as shown by the upper set of red diamonds in
Fig.~\ref{hgl_2dalfa}. The disagreement around $T_{\rm eff}=4\,800$\,K
is most likely also due to differences in the
opacities. They based their opacities on the Atlas6 line opacities whereas
we use the somewhat newer Atlas9 line opacities (in the form of opacity distribution
functions). The difference, as outlined by \citet{kurucz:atm-for-popsynth},
consists of the addition of molecular opacity (hydrides and CN, C$_2$ and TiO)
and improved calculations for the iron-group elements -- all in all a factor
of 34 more molecular, atomic and ionic lines.
\mfigur{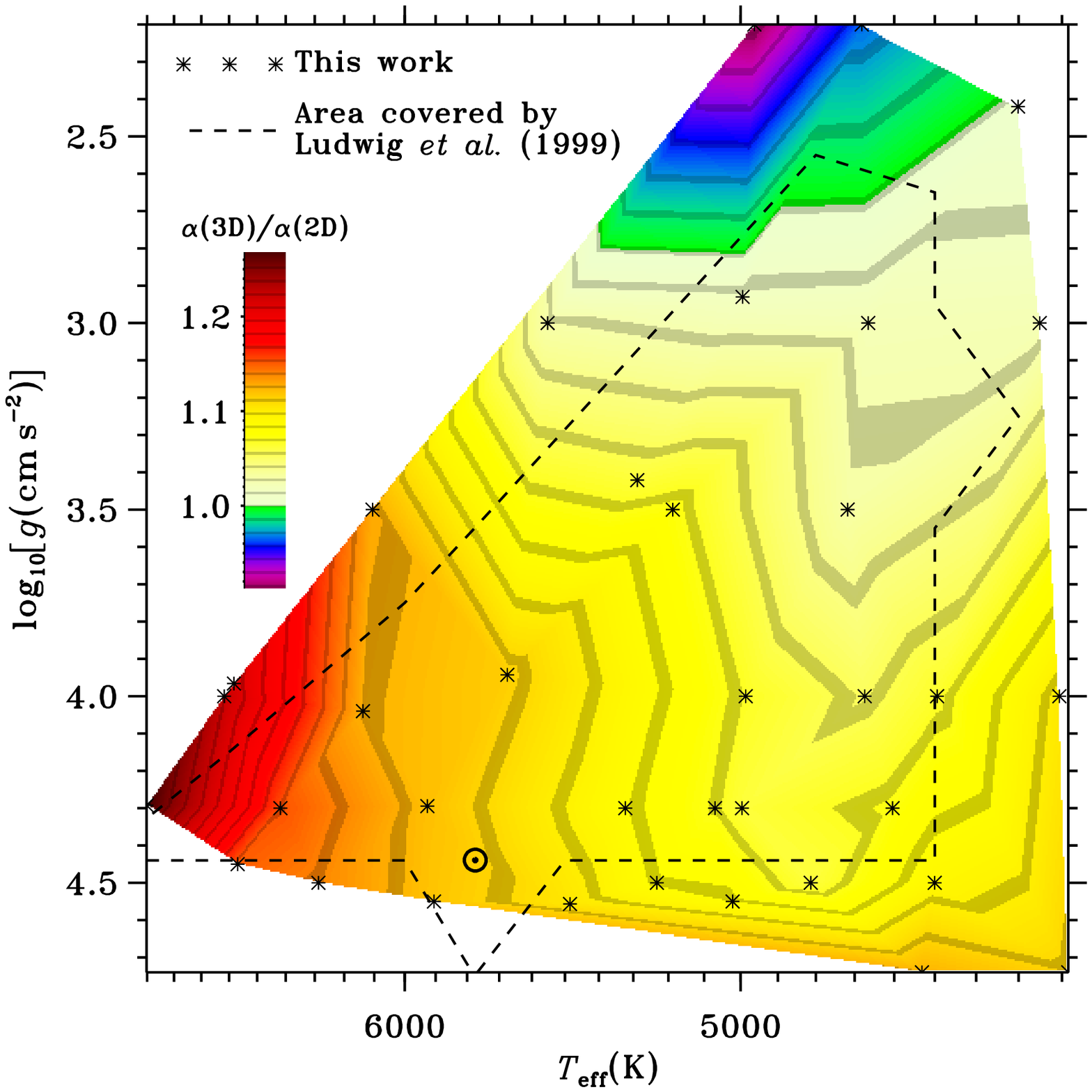}{8.7cm}
	{A comparison between the present $\alpha$ calibration and that by
	 {\protect\citet{ludwig:alfa-cal}} in the $T_{\rm eff}$-/$\log g$-diagram.
     The area covered by their simulations is shown with the dashed outline,
     and outside this, their fit which was used for this plot is an
     extrapolation.}

These opacity changes should affect the hotter stars the least, but they
still have an effect on the solar model -- that was, after all, the
main motivation behind the opacity updates \citep{kur:missolar}.
We therefore suspect that the scaling factor, translating LFS' results from
2D to 3D should be based on the hotter stars.
But there are also other differences between our simulations:
LFS used grey radiative transfer in the bulk of the 58 2D simulations
going into their analysis, adding another systematic difference (also
generally decreasing with $T_{\rm eff}$) between our results.
Furthermore, they used the old pre-helioseismology value of the helium abundance,
$Y=28$\%, which effectively lowers the atmospheric opacity, compared with ours
(the He, being an inert gas, displaces high opacity species, as well as
electron donors for H$^-$ formation).

The ratio between the two calibrations is also shown in an atmospheric HR
diagram in Fig.~\ref{cdhgl_2dalfa}, which also shows the extent of LFS's grid.
From this plot it is clear that most of our simulations have $\alpha$ values 
exceeding those of LFS's 2D simulations, except for giants with $\log g\la 2.5$.
Our warmest giant has $\alpha(3D)/\alpha(2D) = 0.92$ but is also an
extrapolation from LFS's grid. The maximum, for our warmest dwarf which is also
within their grid, is 1.27.

It seems natural to expect that quantities other than $T_{\rm eff}$ and
$g_{\rm surf}$ would be more relevant for describing the efficiency of
convection.
The optical depth at the top of the convection zone, for example, seems much
more relevant and directly related to the issue, or the peak amplitude of the
turbulent to total pressure ratio, or the Mach number as suggested by
\citet{samadi:3DGranSpectr2}. These quantities as well as others we have tested,
do not provide simpler or single-valued functions for $\alpha$, and the scatter
is large than the individual $\sigma\alpha$. We conclude that two independent
variables are necessary to describe $\alpha$, and we naturally choose
$T_{\rm eff}$ and $\log g$.

\subsection{Gravity darkening of slowly rotating stars}
\label{sect:GravDark}

A slowly rotating star can, as a first-order approximation, be described as a
set of 1D stellar models with outer boundary conditions that depend on
co-latitude, $\theta$, as $T_{\rm eff}(\theta)$ and $g_{\rm eff}(\theta) = g -
R\Omega^2\sin\theta$, where $R$ is the radius of the star and $\Omega$ is the
(uniform) rotation rate. A formulation was developed by
\citet{zeipel:SlowRotStar2} for stars in radiative equilibrium, and a main
finding\footnote{Whether his result is
predicated on the assumption that the opacity is of the form
$\kappa_{\rm Ross}=f(pT^{-4})$ (compared with the actual $\kappa_{\rm Ross}
\propto \varrho T^9$), is not clear and is beyond the scope of the
present work.} was that $T_{\rm eff}\propto g^{0.25}$.

Building on this, \citet{lucy:GravDark} extended the formulation to also
apply to convective envelopes,
assuming that the various latitudes are connected in
the deep convection zone by a common adiabat.
The generalisation of \citeauthor{zeipel:SlowRotStar2}'s result is
$T_{\rm eff}\propto g^\beta$, defining the 
\emph{gravity-darkening} exponent $\beta$. In case of convective envelopes
and under the common-adiabat assumption,
this gives rise to the adiabatic gravity-darkening exponent
\be{GravExp}
    \beta_{\rm ad} = \dxdycz{\log T_{\rm eff}}{\log g_{\rm eff}}{{\rm asymp, ad}}\ .
\ee
We have evaluated $\beta_{\rm ad}$ by finding the slopes of the adiabats in
Fig.~\ref{cwSmax}, as shown in Fig.~\ref{cwbeta}. The values we find are
similar to the value, $\beta_{\rm ad}\sim 0.08$, which was originally suggested by
\citet{lucy:GravDark}, and also agrees with the values reported by
LFS although we seem to find a larger range.
\mfigur{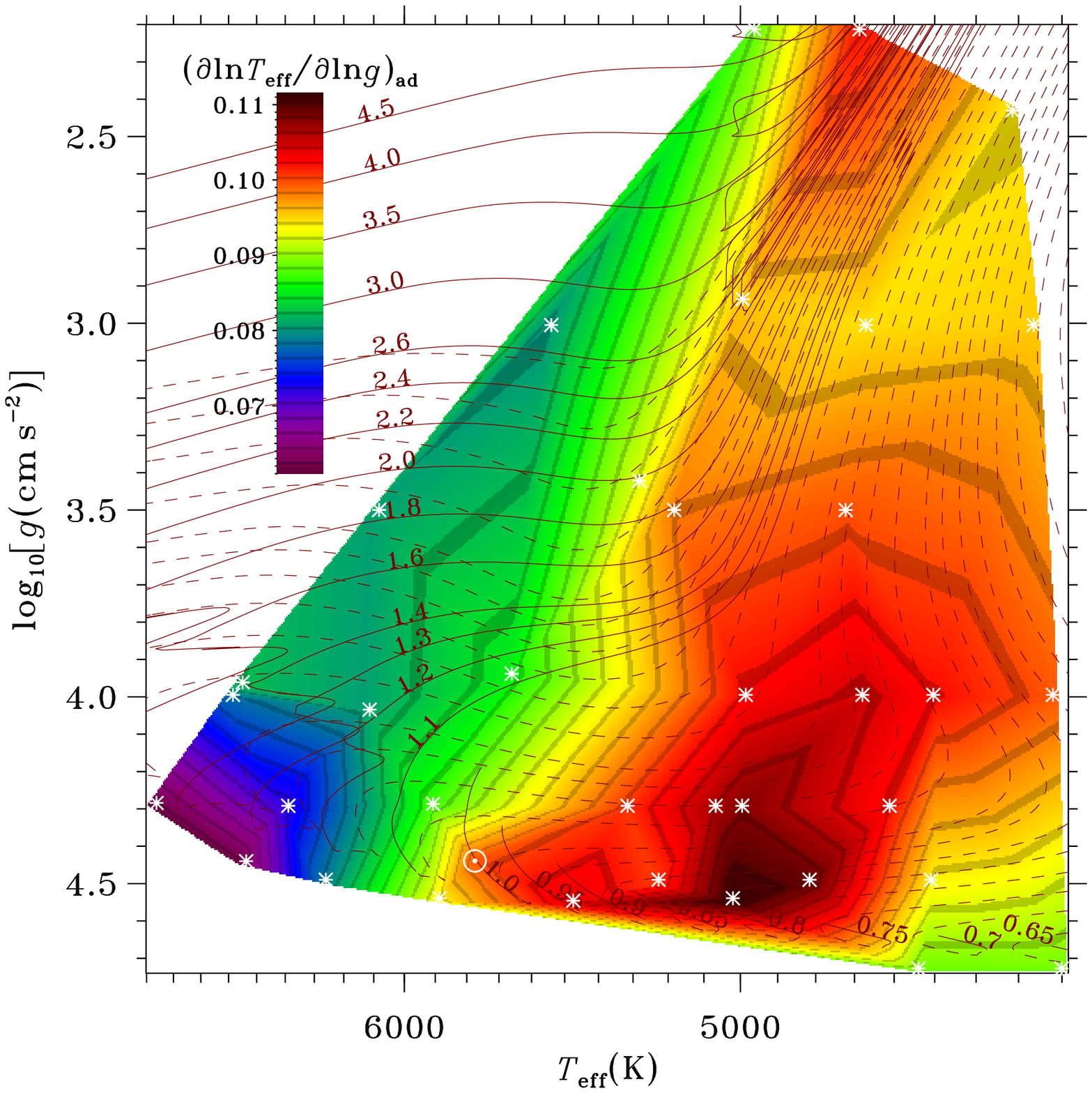}{8.7cm}
    {As Fig.~\ref{cwSmax}, but showing the gravity-darkening exponent, $\beta_{\rm ad}$,
     found from differentiating the asymptotic adiabats of Fig.~\ref{cwSmax}.}

The Sun, with an equatorial rotation rate of 450\,nHz, displays an equatorial
$\log_{10}g_{\rm eff}(\theta=90^\circ)$ that is a mere $8.82\stimes 10^{-6}$\,dex
smaller than at the poles. Making the above assumption of a common adiabat
of the deep convection zone, and using $\beta_{{\rm ad},\odot}=0.100$ from Fig.~\ref{cwbeta}, results
in the equator being an unobservable 12\,mK cooler than the pole (a contrast of $2.1\stimes 10^{-6}$).

On the other hand, it is also quite likely that meridional flows in stars
will build up an entropy gradient with latitude, as suggested by
\citet{miesch:SolarDiffRot}. If we take a latitude-independent $T_{\rm eff}$ as
the other extreme of possibilities, we get an entropy contrast in the solar
convection zone of $1.60\stimes 10^{-6}$. This is similar to the range of
contrasts, 0.95--$4.73\stimes 10^{-6}$, explored by \citet{miesch:SolarDiffRot}
to obtain the observed solar differential rotation in a 3D simulation of the
deep solar convection zone.

Rotational effects are obviously much more important for many other stars.
Also in this regard has NASA's {\it Kepler} mission provided a treasure trove
of observations, with several recent calculations of rotation periods from
spot-modelling of planet candidate host stars by
\citet{walkowicz:KeplerPltHostsRotAge}, \citet{mcquillan:KeplerPltHostRot},
of 12\,000 other F--M {\it Kepler} targets by \citet{nielsen:Kepler12k-rot}
and of 34\,000 {\it Kepler} main-sequence targets cooler than 6\,500\,K by
\citet{mcquillan:34kStarRots} with periods down to 0.2\,days.

For fast rotators, more elaborate models are needed [See, e.g.,
\citet{espinosa-lara:GravDarkFastRot} for analytical 2D work and
\citet{augustson:Fstars1} for 3D hydrodynamics simulations], but the behaviour
of $S_{\max}(T_{\rm eff}, \log g)$ in Fig.~\ref{cwSmax} should still be
relevant for such modelling.

\section{Implications for stellar structure}
\label{sect:StelStruc}


In this section we explore the effects of outer boundary conditions on the 1D
structure models, introduced in Sect.~\ref{sect:env}. We evaluate sensitivities
to changes in $\alpha$, {\Ttau}s and atmospheric opacities in
Sect.~\ref{sect:dCZ} and employ the {\Ttau}s of Paper\,I, together with other
commonly used choices, in Sect.~\ref{sect:Ttaueff}, commenting on the effects
on the depths of convective envelopes.
This analysis does not include calibrations of $\alpha$, but address how global
properties of 1D models react to independent changes to $\kappa_{\rm atm}$,
{\Ttau} and $\alpha$.

The reason for convection zones growing with increasing convective efficiency,
$\alpha$, is found in Sect.~\ref{sect:posddCZda}.

\subsection{The depth of outer convection zones}
\label{sect:dCZ}

The depth of an outer convection zone depends in a complex way on the surface
boundary conditions. With some simplifications, however, a rough idea of the
mechanisms involved can still be obtained. We convert the equation of
hydrostatic equilibrium from the conventional height scale
to an optical depth scale
\be{hydeqtau}
    \dd{p}{\tau} = {g\over\kappa}\ ,
\ee
and integrate inward from $\tau=0$ to get the pressure with depth. For the
present discussion we only need the differential response to changes in the
atmosphere and the precise values of the quantities are immaterial. We
therefore write the pressure in the photosphere
\be{hydeqtauint}
    p_{\rm ph} = {g\over\bar\kappa}\bar\tau\ ,
\ee
as a one-step numerical integration from $p=0$ to $p_{\rm ph}$ and $\tau=0$ to 
$\bar\tau$,
where $\bar\kappa$ and $\bar\tau$ are some appropriate averages over the atmosphere. This results in
a first-order estimate of the effects of changing various parts of the physics
in the atmosphere.

Using some average of the inverse {\Ttau} for $\bar\tau$,
a relation between $T$ and $p$ is obtained. An increase in $T(\tau)$
will decrease $\tau(T)$,
as the {\Ttau} is monotonically increasing, and will therefore
have the same decreasing effect on $p$ as will an increase in the opacity. 
Changes to the {\Ttau} are performed on the generalised Hopf functions,
equation~(\ref{eq:hopf}), which are normalised by the surface flux.

%
We will now assume that an atmospheric pressure change [due to changes to
$\kappa_{\rm atm}$ or $q(\tau)$], represents the same factor of pressure change
throughout the convection zone.
The change in the depth of the convection zone can be derived from the response
to such a pressure change at the bottom of the convection zone. The
Schwarzschild criterion for convection to occur, equation~(\ref{conv-crit}),
is mainly governed by $\nabla_{\rm rad}$, as the adiabatic gradient is very
close to the ideal- and fully-ionised-gas value of
$\nabla_{\rm ad}=\frac{2}{5}$ at the bottom of deep convection zones.

In the deep interior, where the Hopf function, equation~(\ref{eq:hopf}), is constant and the turbulent
pressure is insignificant, the radiative temperature gradient is
\be{nab-rad}
    \nabla_{\rm rad}={3\over 16\sigma}{\kappa F_{\rm tot}p\over gT^4}\ .
\ee
This gradient
will decrease with a decrease in pressure and the bottom of the convection zone
will therefore move outward. Since $\nabla_{\rm rad}$ depends strongly
on temperature and has a steep gradient at the bottom of the convection
zone, the pressure change hardly affects the location of $\nabla_{\rm rad}=
\nabla_{\rm ad}$ on the temperature scale.
The largest effect is therefore due to the (almost unchanged) temperature
at the bottom of the convection zone occurring at a smaller pressure.

This trend is confirmed by the experiments.
We calculated envelope models with small (additive) changes (0.005) to
$\alpha$, $q$ and
the atmospheric opacity, $\ln\kappa_{\rm atm}$
in order to find the differential changes to
the relative depth $d_{\rm cz}$ (in units of stellar radius, $R$)
of the convection zone.
The fundamental parameters, $T_{\rm eff}$, $g_{\rm surf}$ and $M$ (and hence
also $R$ and $L$) were held
constant between the experiments.
Keeping $T_{\rm eff}$ constant effectively anchors the temperature structure
so that the effect of the perturbations in $\alpha$, $q$ and
$\ln\kappa_{\rm atm}$ give similar magnitude responses in $\varrho$ and $p$
(and close to constant in the convective envelope) and an order of magnitude
smaller response in $T$.

\mfigur{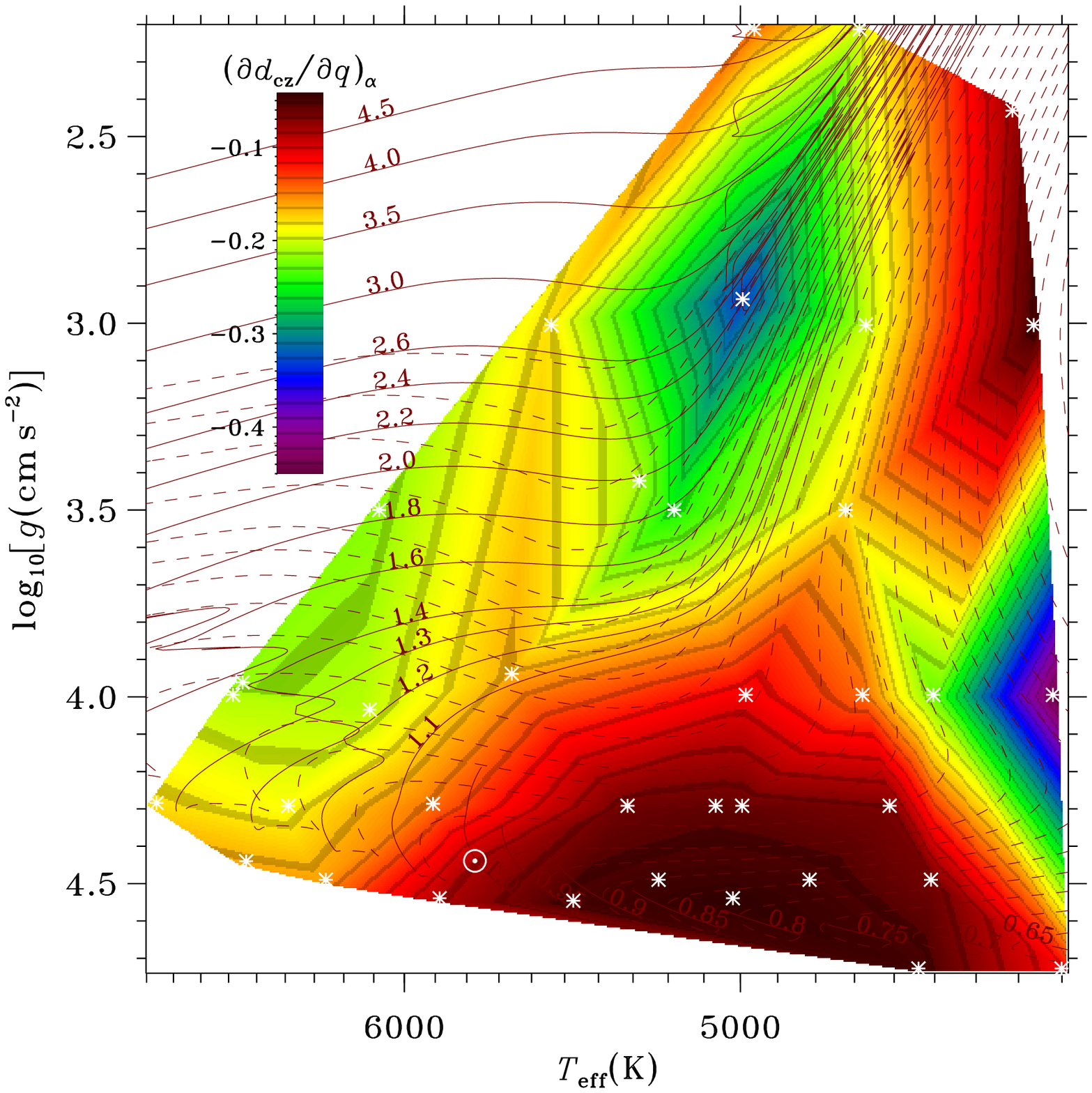}{8.7cm}
    {As Fig.~\ref{cwSmax}, but showing the differential response of the depth
     of the convection zone in units of stellar radii, $R$, to a change in the {\Ttau}, using $\delta q=5\times10^{-3}$.
     Our simulations all display shallower convective envelopes with increasing
     $q$.}
The results of these experiments are presented in
Figs.~\ref{cwddczdq}--\ref{cwddczda}, in the form of changes to the depth of
the convective envelope, as function of the atmospheric parameters $T_{\rm eff}$
and $\log g$.
Our experiments shows that both $\upartial d_{\rm cz}/\upartial q$
in Fig.~\ref{cwddczdq} and $\upartial d_{\rm cz}/\upartial\ln\kappa_{\rm atm}$
in Fig.~\ref{cwddczdk} are negative and have a similar
functional form so that $\upartial d_{\rm cz}/\upartial q \simeq
0.4\stimes \upartial d_{\rm cz}/\upartial\ln\kappa_{\rm atm}$. The depth of
the convection zone
increases with increasing $\alpha$, as expected (see Sect.~\ref{sect:posddCZda}),
and we have approximately $\upartial d_{\rm cz}/\upartial\alpha \simeq
-0.7\stimes \upartial d_{\rm cz}/\upartial\ln\kappa_{\rm atm}$. 

\mfigur{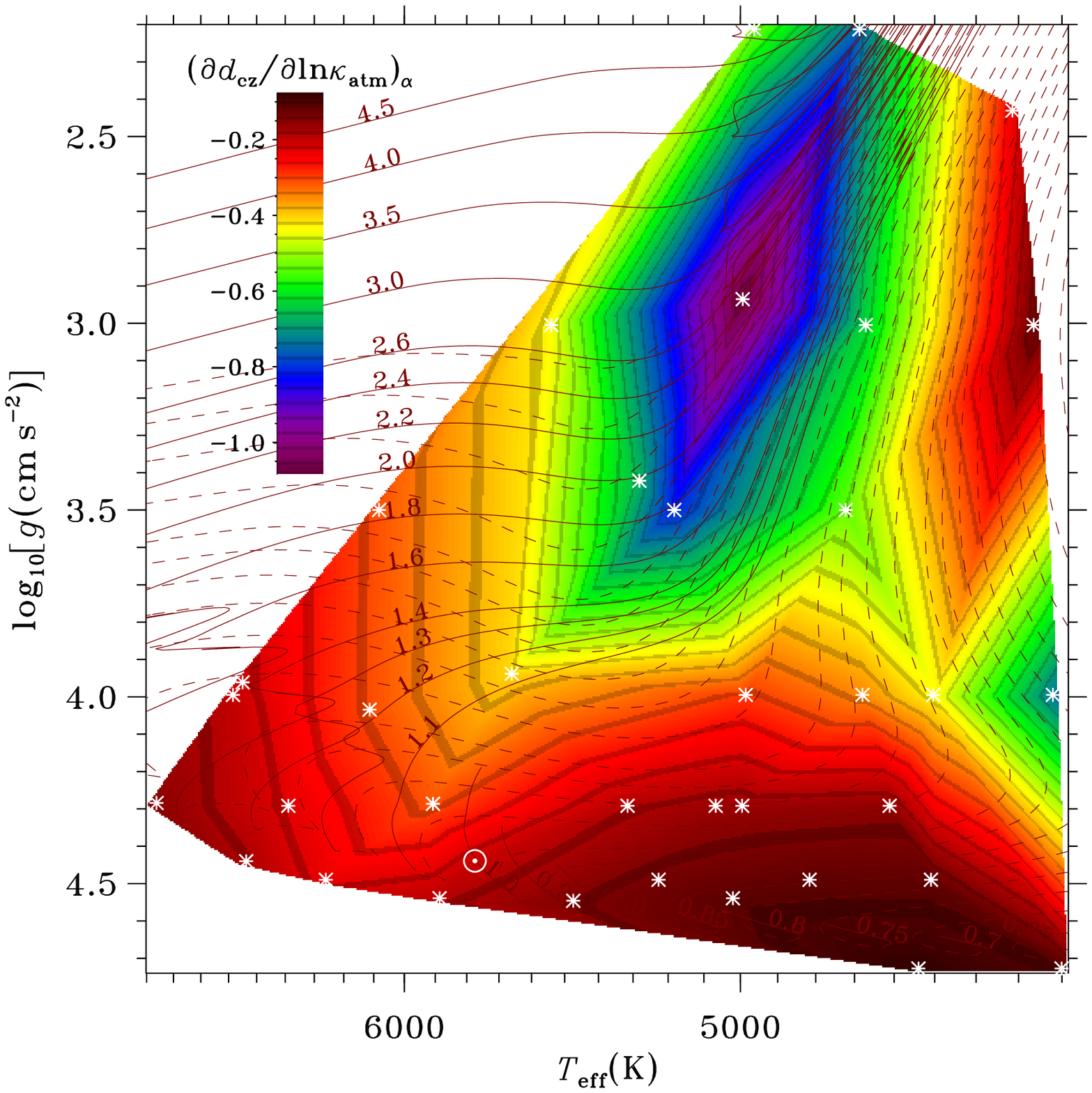}{8.7cm}
    {As Fig.~\ref{cwSmax}, but showing the
     differential response of the depth of the convection zone, to a
     change in the atmospheric opacity, $\delta\ln\kappa_{\rm atm}=5\times10^{-3}$.
     All our simulations have shallower convective envelopes with increasing
     $\ln\kappa_{\rm atm}$.}
%
%

The convective flux in the MLT formulation may be written as
\be{MLT-Fconv}
    F_{\rm conv} \propto \alpha^2 {p_{\rm g}^2\over T^{3/2}}
                   (\nabla - \nabla^\prime)^{3/2} ,
\ee
where $\nabla^\prime$ is the temperature gradient in the upflowing convective
elements and $\nabla$ is the average of the gradient between the
up- and downflowing elements. Note that this is all in the 1D mixing-length
picture, so the average gradient $\nabla$, is not the average of a 3D quantity
which would have been denoted by $\langle\nabla\rangle$.
The difference between $\nabla$ and $\nabla^\prime$ is
almost equal to the superadiabatic gradient, $\nabla - \nabla_{\rm ad}$.
Based on equation~(\ref{MLT-Fconv}), an increase in temperature and/or a decrease
in pressure (brought about by changes to the {\Ttau} or the atmospheric
opacity) will therefore be
accompanied by an increase in $\nabla - \nabla_{\rm ad}$ in order to
maintain the total flux (and the fixed $T_{\rm eff}$ of the model).
This increase of $\nabla - \nabla_{\rm ad}$, corresponding to a
decrease of the efficiency of the convection, will lead to a smaller
convection zone. The effect can be counteracted by increasing $\alpha$, which
increases the convective efficiency by increasing the distance travelled by convective
elements. An increase in the efficiency of convection will enlarge the
convection zone, as seen from Fig.~\ref{cwddczda} 
\mfigur{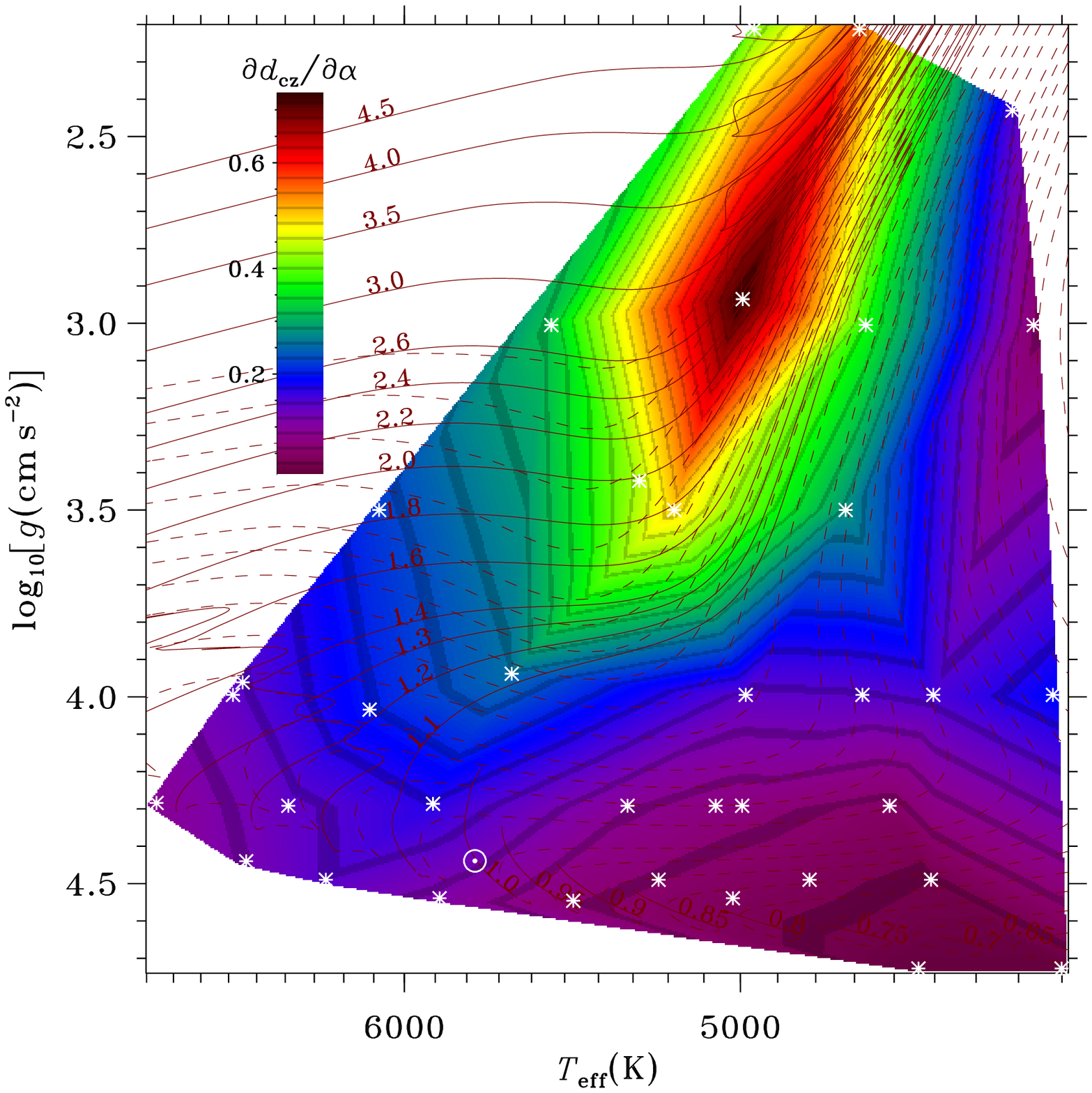}{8.7cm}
    {As Fig.~\ref{cwSmax}, but showing the
     differential response of the depth of the convection zone, to a
     change in the convection parameter, $\delta\alpha=5\times10^{-3}$.
     All our simulations display a deepening of their convective envelopes
     with increasing $\alpha$, as expected.}
(see also Sect.~\ref{sect:posddCZda} below).

As far as global
observables are concerned, uncertainties in the atmospheric opacities,
line-blocking and mismatches between the $\tau$ scale and the grey opacity may
be hidden in $\alpha$ {\em together} with pure convection effects. This is one
reason that so many values for the solar $\alpha$ can be found
in the literature (another reason being the lack of consensus on the values for
the auxiliary MLT parameters).
As illustrated by Figs.~\ref{cwddczdq}--\ref{cwddczda}, such flaws in the
treatment of the
photosphere will not have the same effect on all stars and an incorrect
differential behaviour would be expected from such a solar calibration,
masking real convection effects ({\ie}, an $\alpha$ that varies with atmospheric
parameters).

Since it is customary to use a solar-calibrated $\alpha$ for computing stellar
models, we have also computed a set of envelopes with constant $\alpha=
\alpha_\odot$, but still using the individual {\Ttau}s. The effect of
this is to overestimate the fractional depth of the convection zone of giants
by 4--8\% and underestimate it for cool dwarfs by about 1\%, as shown in Fig.~\ref{cdasun}. This effect is somewhat counteracted when also keeping the
{\Ttau}s constant (scaled solar). The result of Fig.~\ref{cdasun} is then
further changed by a reduction of $d_{\rm cz}$ along the warm edge of our grid
by 1.7\% and an increase of $d_{\rm cz}$ by about 5\% along the cool edge.
\mfigur{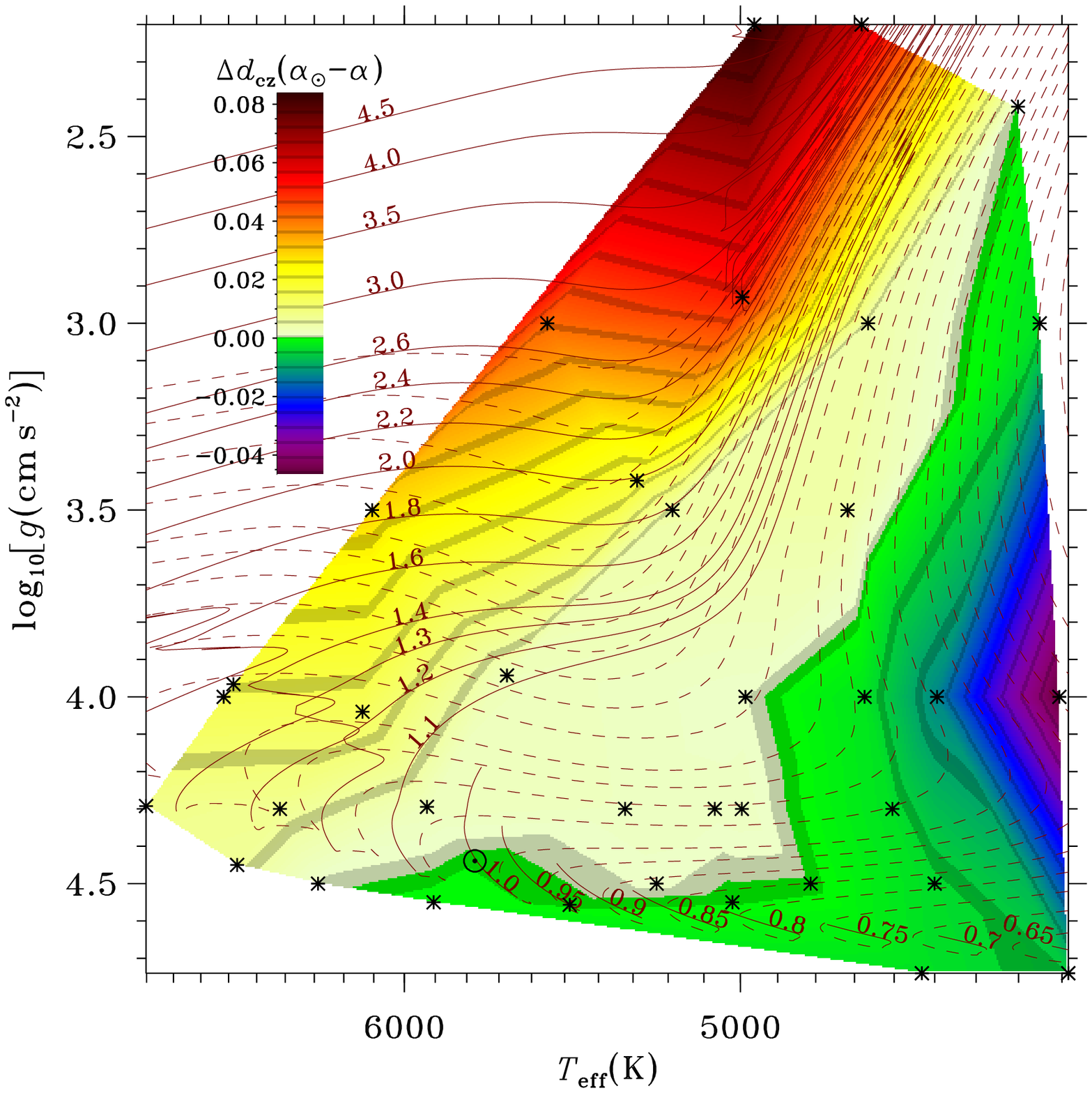}{8.7cm}
    {As Fig.~\ref{cwSmax}, but showing differences in the depth of the
     convective envelope between models using solar-calibrated and individually
     calibrated values of $\alpha$, in the sense that warm giants with
     $\alpha_\odot$ have deeper convective envelopes.}

\subsection{Effects of changing the $T(\tau)$ relations}
\label{sect:Ttaueff}

Since there currently are a number of schemes for using {\Ttau}s in stellar
models, we find it worthwhile to explore the consequences of a few of these.
We have kept the opacity and the mixing length, $\alpha$, unchanged in the
experiments below.

As mentioned earlier, one of the first solar {\Ttau}s to be published
\citep{krishna-swamy:lines-K-dwarfs}, is still in wide-spread use in stellar
models. Fig.~\ref{cwdqKS} shows the effect of using this {\Ttau} (scaled to
the stellar $T_{\rm eff}$) instead of the individual ones found from the
simulations.
\mfigur{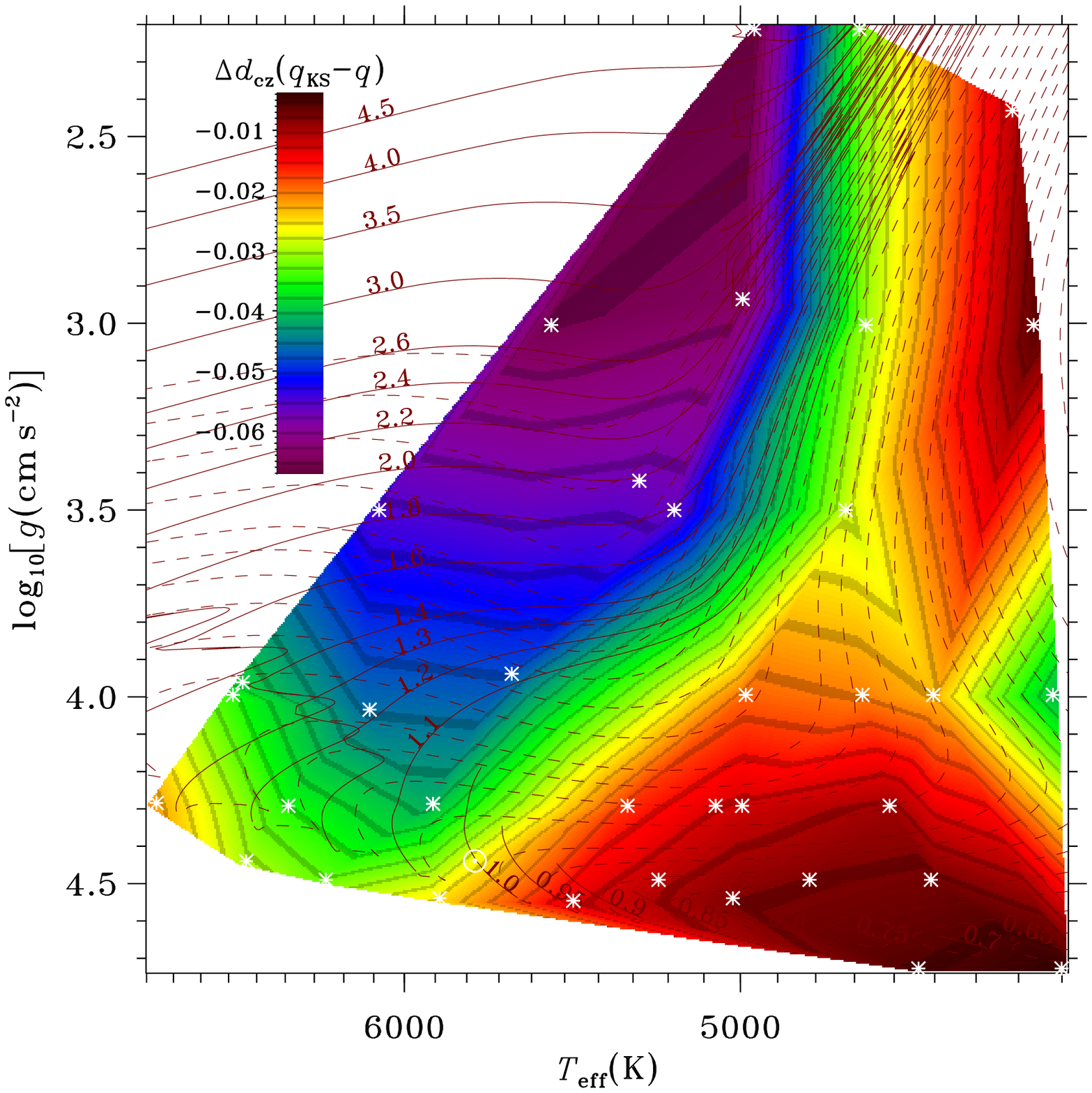}{8.7cm}
    {Envelope models employing the
    {\protect\citet{krishna-swamy:lines-K-dwarfs}} {\Ttau} instead of the
     individual ones derived from the simulations have a systematically
     shallower convection zone. Also compare with Fig.~\ref{cwddczdq}.
     See Fig.~\ref{cwSmax} for further explanation of the plot.}
The effect is largest for a $\sim$4.5\,M$_\odot$ star on the horizontal branch (HB), or a slightly more massive star approaching the Hayashi line after crossing the Hertzsprung gap,
which has its convection zone decreased from 28\% to 21\% of the stars radius,
from employing the \citeauthor{krishna-swamy:lines-K-dwarfs}
{\Ttau}. The
smallest effect (apart from the fully convective No.~4) is found in the cool
dwarf corner of the HR diagram (simulation No.\ 36 of Table~\ref{starlist}), where the
convection zone depth is only decreased from 33.3\% to 33.0\%. The solar convection zone shrinks
from 27.9\% to 25.2\% of R$_\odot$, from this change.

Another common approximation is to use a scaled solar {\Ttau}, albeit a more
modern one. Using our solar {\Ttau} from the convection simulations, we find
a change somewhat similar to that depicted in Fig.~\ref{cwdqKS}, although with
half the range and the solar location defining the zero-point; The cool dwarfs
get larger convection zones by 0.04 of their radii and the 4.5\,M$_\odot$ HB star
has its convection zone decreased by 0.018 of
its radius.

Many semi-empirical solar atmosphere models do not include
the Rosseland optical depth scale, but are only given on the monochromatic 5000\,{\AA}, $\tau_{5000}$-scale. On the other hand, opacities for stellar structure
calculations only include Rosseland averages, but not the 5000\,{\AA} opacity.
When trying to combine $T(\tau_{5000})$ and $\kappa_{\rm Ross}$ in the
atmosphere of a stellar structure
calculation, an obvious inconsistency emerges, which has been largely ignored.
Fig.~\ref{cwdq5000} shows to what extent this is a good approximation.
\mfigur{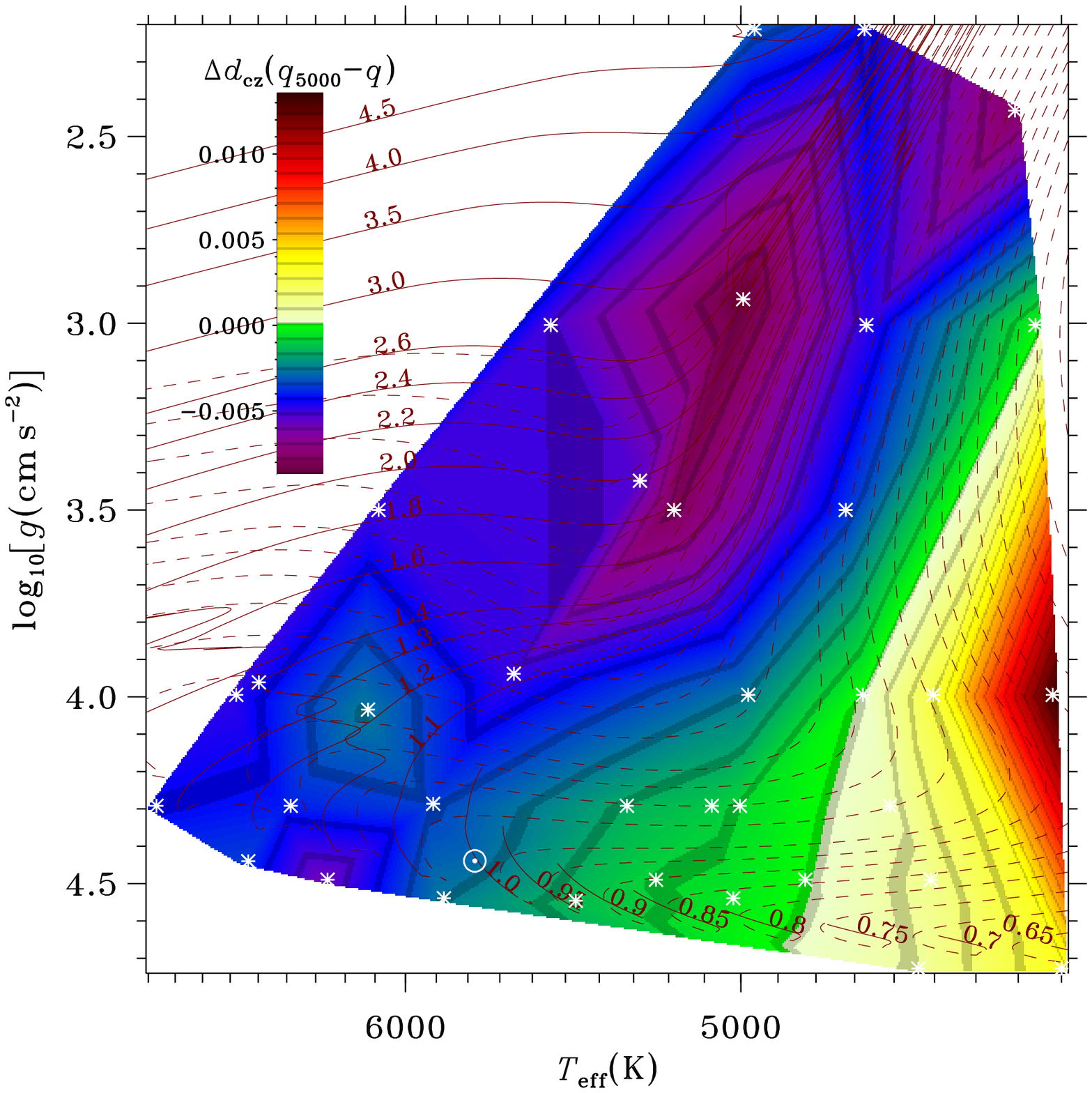}{8.7cm}
    {Envelope models employing {\Ttau}s on the monochromatic $\tau_{5000}$-scale
     in combination with the usual Rosseland opacities have systematically
     shallower convection zones, except for the coolest part of the grid where
     the convective envelopes deepen. Also compare with Fig.~\ref{cwddczdq}.
     See Fig.~\ref{cwSmax} for further explanation of plot.}
It shows how the depth of convection zones of 1D envelope models changes when
using $T(\tau_{5000})$ relations from the simulations together with Rosseland opacities.
We see that the change is fairly small, mostly well below 1\% of the stellar radius, with models more massive than the Sun having deeper convective envelopes
when applying a consistent set of {\Ttau}s and opacities. Our solar case has
a convection zone deeper by 0.3\% of R$_\odot$ with the consistent combination. The one
simulation with a fairly large change, 1.4\% of its radius, also has the
largest $\kappa_{5000}/\kappa_{\rm Ross}$-ratio in the photosphere.
This ratio only exceeds unity in the photospheres of the stars along the
low-$T_{\rm eff}$ border of our grid. This ratio \emph{does} exceed unity 
for all the other simulations at some point higher in their atmospheres.

We recommend to use modern {\Ttau} based on modern opacities, and if possible,
using those same opacities in the atmospheres of the stellar structure models.
This is both to ensure consistency and in order to try to separate 
radiative and convective effects. This makes it possible to have the correct
behaviour for stars other than the Sun, which still provides the strongest
observational constraints for an $\alpha$ calibration.

From the above analysis, we have seen how the depth of the convective envelope
depends on three parameters of the atmosphere: the mixing length, $\alpha$,
the {\Ttau} and the opacity. Conversely, it is also clear
that an inadequate knowledge of {\Ttau} and atmospheric opacity can be absorbed
into $\alpha$, so that different MLT parameters can
result in the same overall stellar structure ({\ie}, result in the deep
convective envelope lying on the same adiabat). This is the most likely reason
for the variation in the solar-calibrated $\alpha_\odot$ in the literature.
Consistent treatment of the atmosphere will hopefully tighten the range of
$\alpha_\odot$.

\subsection{Calibrating $\alpha$ with fixed $T(\tau)$ relation}
\label{sect:constTtau-alpha}

In many cases stellar models are computed using scaled solar {\Ttau}s; this
includes many of the $\alpha$ calibrations against observations discussed
below (see Sect.\ \ref{sect:OtherCals}). We
therefore performed such a calibration, based on a fit to the quiet-Sun
stratification of the semi-empirical atmosphere model of \citet{VAL-III},
referred to as VAL-C. 
\mfigur{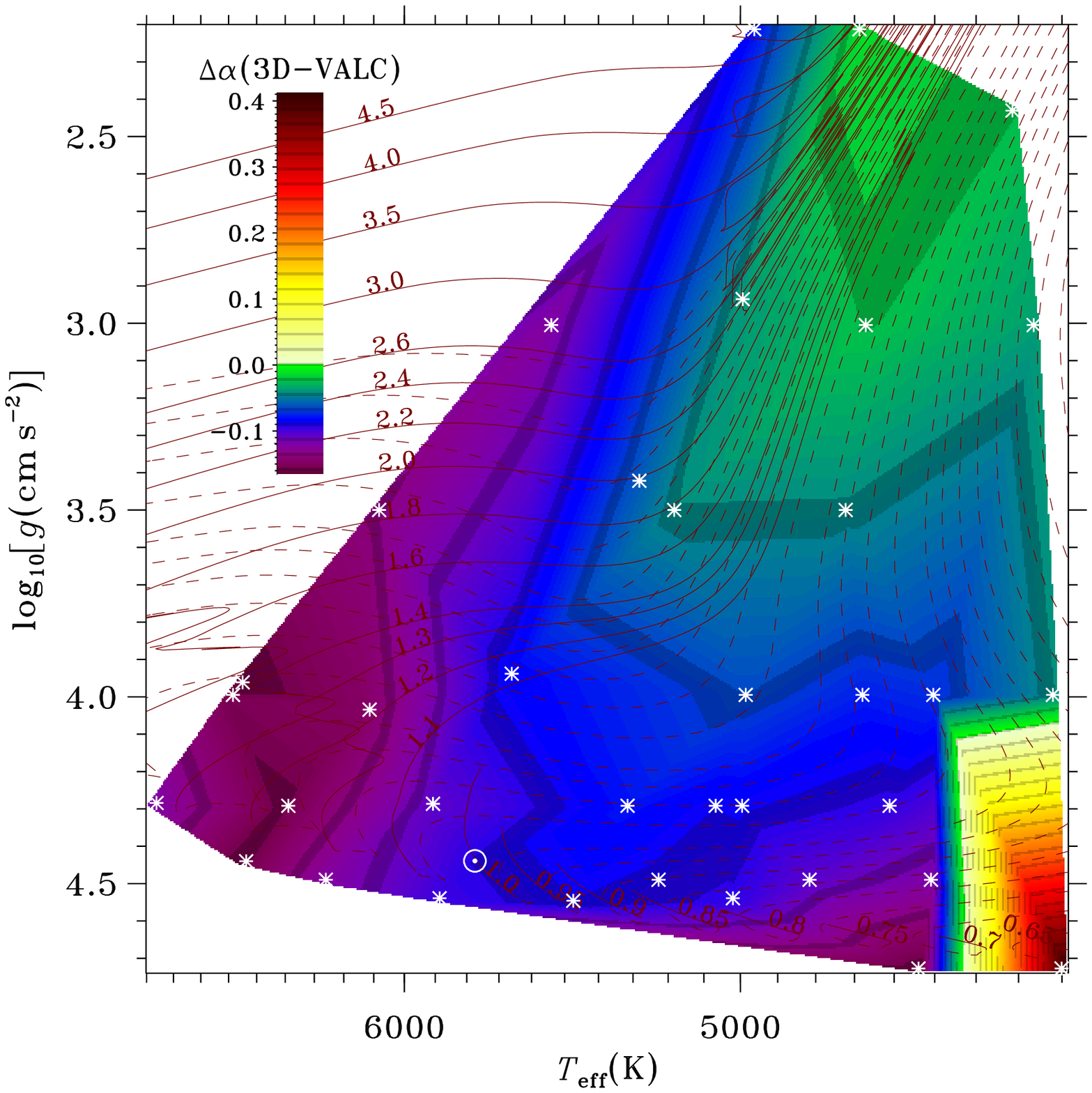}{8.7cm}
    {A calibration of MLT $\alpha$ based on 1D envelope models using scaled
     solar {\Ttau}s from the semi-empirical VAL-C solar atmosphere model
     {\protect\citet{VAL-III}}. We show the difference in $\alpha$ from this
     calibration, and the one shown in Fig.~\ref{cwalfa} based on individual
     {\Ttau}s of the convection simulations. The singular positive value for
     our coolest dwarf is discussed in Sect.\ \ref{sect:constTtau-alpha}.}
This choice of {\Ttau} is the default in, e.g., the ASTEC stellar structure code
\citep{jcd:ASTEC} and the closely related stellar envelope code employed here.
ASTEC is also part of the Asteroseismic Modeling Portal (AMP) by
\citet{metcalfe:AMP+Kepler}. AMP was used for the the seismic
$\alpha$ calibrations of \citet{savita:Kpl22sols},
\citet{bonaca:KeplerAlpha} and \citet{metcalfe:AMP42Kepler}. For this experiment only, we use the envelope code
with its default choice of {\Ttau}.

In Fig.~\ref{cwdadqVAL} we show the differences in $\alpha$ between this
calibration with fixed {\Ttau}, and our full calibration with individual
{\Ttau}s from the simulations (Sect.\ \ref{sect:results} and Fig.~\ref{cwalfa}).
In both cases the calibrations are performed as outlined in Sect.\
\ref{sect:env-match}, the only difference being the choice of {\Ttau}.
We note that with the
VAL-C atmosphere, $\alpha$ calibrates to systematically larger values than in
the full calibration, except for our coolest dwarf simulation (No.\ 36 of Table
\ref{starlist}). For that case $\alpha({\rm VALC})$ is 0.41 smaller.
This is a robust result. We have traced
this change of sign and magnitude of $\alpha$'s sensitivity to the {\Ttau},
to a an extended overlap of the photosphere and the top of the convection zone
in the envelope model. The convective flux increases from 3.2\% to 28\% of
the total flux, at $\tau=1$ between the two coolest dwarf simulations in our grid.
Constructing 10 envelope models, with parameters linearly interpolated between
the two calibrated models, we find a continuous, but rapid increase of the
photospheric convective flux fraction towards lower $T_{\rm eff}$. At $\tau=2$ the increase
is from 6.7\% to 43\%. The consequence is that the {\Ttau} very directly
determines the adiabat of the convection zone of the coolest dwarf, whereas
the warmer models have the structure effects of the {\Ttau} diminish with
depth, before appreciable convection is reached. We have further attributed the
outward migration of the convection zone to the effect of H$_2$
dissociation on $\nabla_{\rm ad}$, which for this cool dwarf model is suppressed
from its fully ionised value of 2/5 to 0.095 centred at $\log\tau=-2.8$ and
a value of 0.23 at $\tau=1$. The outward decrease in $\nabla_{\rm ad}$
induces a shoulder in $F_{\rm conv}$, increasing in amplitude as the H$_2$
feature in $\nabla_{\rm ad}$ moves inward with decreasing $T_{\rm eff}$.
This extension of convection into the photosphere by H$_2$ dissociation, was
also found by \citet{aake:osc-conv}.

We also see this phenomenon in our 3D simulation, with similar behaviour
of $\nabla_{\rm ad}$ and a corresponding shoulder on $F_{\rm conv}$. In 3D the
feature is smoothed, compared with the 1D model, by the convective fluctuations.
The effect is therefore not due to the atmospheric simplifications of 1D models;
rather its physical reality is supported by our 3D simulations.

The effect of the difference in {\Ttau} is to increase the overall density of
the 1D envelope by about 0.1\,dex, translating into the large difference in
$\alpha$. For all the other cases, the resulting change in interior density is
small and negative. We have not yet found the underlying reason for this
difference. For our solar case, $\alpha({\rm VALC})$ is 0.09 larger than the
calibration with its own 3D {\Ttau}. The difference is a minimal 0.013 for our
$T_{\rm eff}=4\,962$\,K giant (simulation No.\ 2 of Table~\ref{starlist}) and
largest for the simulations with $T_{\rm eff}$ around 6\,400\,K, as well as for
the next coolest dwarf (simulation No.\ 37 of Table~\ref{starlist}).


\subsection{Is $\upartial d_{\rm cz}/\upartial\alpha$ always positive?}
\label{sect:posddCZda}

An analytical analysis of $\upartial d_{\rm cz}/\upartial\alpha$
was carried out by \citet{jcd:score96} (hereafter \hbox{C-D97}).
His analysis was concerned with a highly simplified but instructive model:
approximating the convective envelope with a mass-less (assuming all the
stellar mass resides in the radiative interior) polytrope of index $\gamma$,
which relates pressure, $p$, to density, $\varrho$, through $p=K\varrho^\gamma$.
Further
assuming hydrostatic equilibrium and a fully ionised perfect gas, he obtained
the differential of the stellar radius, $R$, (\hbox{C-D97} equation~[9])
\be{dR}
    \upartial R = -H_p\left(\frac{R}{r_{\rm cz}}\right)^2\upartial\ln p_{\rm cz}
               + \frac{R^2d_{\rm cz}}{r_{\rm cz}}\left[
                 \frac{\upartial\ln K}{\gamma}
                 + \frac{\gamma-1}{\gamma}\upartial\ln p_{\rm cz}\right]
\ee
where $r_{\rm cz}$
\footnote{Note that \hbox{C-D97}
used the symbol $d_{\rm cz}$ for the absolute depth of the convection zone,
whereas we use it for the relative depth. Compared with his equations, all
occurrences of $d_{\rm cz}$ are therefore multiplied by $R$ ($r_{\rm cz}$
is unchanged).}
is the radius of the bottom of the convection zone and $H_p$
is the pressure scale height at $r_{\rm cz}$.
Using \hbox{C-D97's} equation~[6] we define $C_1$
\be{C1}
    C_1\equiv-\frac{4-\kappa_T}{(4-\kappa_T)(\gamma-1)-\gamma(\kappa_p+1)}
        \simeq \dxdy{\ln p_{\rm cz}}{\ln K}\ ,
\ee
which 
contains the temperature and pressure derivatives of the Rosseland opacity,
\be{kapder}
    \kappa_T \equiv \left({\upartial \ln\kappa \over \upartial\ln T}\right)_p\ ,
\qquad
    \kappa_p \equiv \left({\upartial \ln\kappa \over \upartial\ln p}\right)_T\ ,
\ee
at $r_{\rm cz}$.
With equation~(\ref{C1}) we can eliminate the pressure, $p_{\rm cz}$, in
equation~(\ref{dR}) and using \hbox{C-D97's} equation~[15]
\be{dlnKdlna}
    \dxdy{\ln K}{\ln\alpha} \simeq \frac{2\Delta s}{c_p}\, 
\ee
where $c_p$ is the specific heat at constant pressure at $r_{\rm cz}$,
we arrive at
\be{dRdlna}
        \dxdy{R}{\ln\alpha} =
        \left\{C_1H_p\left(\frac{R}{r_{\rm cz}}\right)^2
        - \frac{R^2d_{\rm cz}}{r_{\rm cz}}\left[C_1+\frac{1-C_1}{\gamma}\right]
        \right\}\frac{2\Delta s}{c_p}\ .
\ee
$\Delta s$ is the change of specific entropy from the entropy minimum at the top of the
convection zone, and down to the adiabatic part, integrating over the peak in
the superadiabatic gradient near the surface. This entropy change is
positive according to the Schwarzschild criterion for convective instability,
equation~(\ref{conv-crit}).

For the location of the bottom of the convection zone, \hbox{C-D97} found
\be{drczdlna}
        \dxdy{r_{\rm cz}}{\ln\alpha} = C_1H_p\frac{2\Delta s}{c_p}\ .
\ee
Combining equations~(\ref{dRdlna}) and (\ref{drczdlna}) with $\delta(Rd_{\rm cz}) =
\delta R - \delta r_{\rm cz}$, and
\be{deldcz}
        \delta d_{\rm cz} = \frac{\delta(Rd_{\rm cz})}{R} -
                                                d_{\rm cz}\frac{\delta R}{R}\ ,
\ee
we finally obtain
\bea{ddCZdlna}
        \dxdy{d_{\rm cz}}{\ln\alpha} &=& \left\{
                \frac{C_1 H_p}{r_{\rm cz}} - C_1-\frac{1-C_1}{\gamma}
                \right\}d_{\rm cz}\frac{2\Delta s}{c_p}\ .
\eea
In order for this to be positive, we therefore require 
the curly bracket to be positive. This can be recast into the
surprisingly simple inequality
\be{posddczda}
        \frac{r_{\rm cz}}{H_p} > \frac{4-\kappa_T}{\kappa_p+1}
        \qquad{\rm for} \quad C_1 < 0\ ,
\ee
(assuming that $4-\kappa_T > 0$)
and the opposite inequality for $C_1 > 0$.

Under which circumstances do we have $C_1 < 0$? From equation~(\ref{C1}) we find
that this is the case when
\be{C1neg}
        \gamma > \frac{4-\kappa_T}{4-\kappa_T-(\kappa_p+1)} \; .
\ee
At the bottom of convective envelopes this is often fulfilled since
$\kappa_T$ is negative and of large absolute value
and $\kappa_p$ is positive and small, so that
the right-hand side of equation~(\ref{C1neg}) will not be much larger than 1.
In the solar case, with $\kappa_T=-3.61$ and $\kappa_p=0.58$ we get 1.22,
which indeed is smaller than the fully ionised, ideal gas value of $\gamma=5/3$.

The bottom of a convective envelope occurs where the radiative temperature
gradient drops below the adiabatic temperature gradient
[{\cf} equations~(\ref{conv-crit}) and (\ref{nab-rad})].
Assuming that the convective
envelope is {\em not} deep enough to reach into the core
(in which case the assumptions of mass-less envelope and constant luminosity
would break down),
the
drop in $\nabla_{\rm rad}$ will be due to a decrease in opacity. We therefore
expect a large and negative $\kappa_T$. The pressure dependence of the opacity
is generally much weaker than the temperature dependence, and it is in general
positive.
The criterion for the depth of a convective envelope to increase with
$\alpha$, will therefore in general be that of equation~(\ref{posddczda}), which
is fulfilled for all the stars considered in the present paper. In particular,
for the Sun the above analysis results in
$\upartial d_{\rm cz}/\upartial\alpha=0.16$.
Despite the differences in the two methods; analytic versus numerical and
simplified full stellar model versus detailed, but truncated envelope models,
this value is close to our result, $\upartial d_{\rm cz}/\upartial\alpha=0.12$.

\section{Observational constraints}
\label{sect:ObsCmp}
\subsection{Depth of the solar convection zone}
\label{sect:solardCZ}

One of the simulations in our grid, No.\ 30, corresponds to the Sun,
and we have carefully adjusted the
entropy of the inflowing gas (a constant) to obtain an effective temperature
of $5\,774\pm 15$\,K, in agreement with that derived from total solar irradiance
(TSI) observations: $T_{{\rm eff},\odot}=5\,777\pm 2.5$\,K, \citep{willson:solar-Irr}.
A recent, but contentious, reassessment by \citet{kopp:SolarIrrAgree} of a number of space-based,
TSI measurements, find a significantly lower
quiet-Sun TSI of $S_0=(1.3608\pm 0.0005)\stimes 10^6$\,W\,m$^{-2}$, corresponding
to $T_{{\rm eff},\odot}=5\,770.35\pm 0.15$\,K. This is, however, well inside the
RMS-scatter of our solar simulation. The composition of this simulation is
$X=73.70$\,\% and $Z=1.800$\,\%, as for the rest of the grid, and as detailed in
Sect.\ \ref{sect:sims}. This is very close to the $X=73.73$\% and $Z=1.806$\%
composition of the present day convection zone of model S
\citep{GONG-Sci:sol-mod}.

Matching this simulation to an envelope-model gives
$\alpha=1.76\pm 0.03$, $\beta=0.81\pm 0.06$
and a depth of the solar convection zone, 
$d_{\rm cz}=0.2791\pm 0.0009\,$R$_\odot$. This is within a mere 2.5$\sigma$
of the value inferred from inversion of helioseismic observations:
$d_{\rm cz}=0.287\pm 0.003\,$R$_\odot$ \citep{jcd:dCZ} and
$d_{\rm cz}=0.287\pm 0.001\,$R$_\odot$ \citep{basu:dCZ}.
It is interesting to note that the $d_{\rm cz}$ that results from our calibrated
$\alpha$ is only a little deeper than the $d_{\rm cz}=0.276\,$R$_\odot$ found
by, e.g., \citet{serenelli:SunSeismAGSS09}, from matching L$_\odot$ and
R$_\odot$ of models based on the \citet{AGSS2009} abundances.
The uncertainties that we quote for our results are the RMS scatter resulting
from performing
the full fitting of {\Ttau}s and envelope-matching for the
individual time-steps of the relaxed and horizontally averaged simulation. No attempt at accounting for
systematic effects has been carried out here.
%
%

As indicated below equation~(\ref{alpha}), there are two more parameters to standard
MLT: $\Phi$ and $\eta$. These MLT-parameters and $\alpha$ are not linearly
independent and we therefore limit ourselves to add $\eta$ to
our discussion, keeping $\Phi=2$.
As stated earlier, the $\alpha$ calibration presented here, does not reproduce
the atmospheric structure of the 3D simulations, it is rather constructed to
reproduce the structure inside the chosen matching point. That means the extra
free parameter of MLT can be used for matching some other feature of the 3D
simulations. We choose the height of the super-adiabatic peak, and the
amplitude of super adiabaticity at the matching point, as two illustrative
examples.

Fitting $\eta$ with respect to the height of the super-adiabatic peak we get
$\alpha=1.84$, $\beta=0.79$ and $\eta=0.0749$, resulting
in $d_{\rm cz}=0.2792$\,R$_\odot$, also about 2.5$\sigma$
shallower than inferred from helioseismology.
The peak in the super-adiabatic gradient is increased from 0.554
in our standard calibration, to 0.695 with this new value of $\eta$.

If on the other hand we adjust the form factor, $\eta$, so as to obtain the
same $\nabla$ at the matching point, then we get
$\alpha=3.61$, $\beta=0.50$ and $\eta=6.41\times 10^{-4}$,
and a $d_{\rm cz}=0.2816$\,R$_\odot$,
1.7$\sigma$ shallower than the helioseismic result.
However, the peak of the super-adiabatic gradient becomes nonphysically
large, reaching a value of 2.154, about 100\,km below the photosphere. This
is 3.1 times larger than what we find in the solar simulation and even more than twice
as large as the super-adiabatic gradient averaged over only the upflows in the
simulation.

That $T$, $\varrho$ and $\nabla$ cannot be simultaneously matched at a common
pressure-point (with plausible parameters), indicates that the MLT formulation
converges slowly, if ever, towards the super-adiabatic gradient,
$\nabla-\nabla_{\rm ad}$, of a real convective envelope. This might be
due to the neglect of kinetic-flux in the MLT formulation, as discussed in
Sect.\ \ref{sect:MLT-3D}.

Notice that the depth of the solar convection zone, as found above,
results from \emph{ab initio} calculations, from the EOS
and opacity calculations, to the RHD simulations. Apart from the defining
parameters (surface gravity, entropy of the inflows at the bottom, and the
composition), the adjustable parameters
that enter the simulations are the resolution, the viscosity 
coefficients, and the size of the time step relative to the Courant time.
These are tuned to resolve the thermal boundary layer at $\tau = 1$ 
and the convective structures, to minimize numerical diffusion while avoiding
numerical noise, and to minimize the computing time against accuracy.
None of these are adjusted to fit solar observations and these parameters are
therefore not ``adjustable parameters'' in the conventional sense. In
particular, no parameters have been adjusted to obtain a certain atmospheric
entropy jump, or by implication, a certain mixing length, $\alpha$.
The close agreement with helioseismology is therefore very encouraging.

We also note that the $\alpha$ calibration is insensitive to the new abundances
by \citet{AGSS2009}, since these differ by having lower C, N and O abundances
than what we employ, whereas the Fe abundance is unchanged. The latter greatly
affects the solar atmosphere, but C, N and O provide little opacity here,
and have little effect on the solar surface layers. The calibration is
performed entirely on quantities that are minimally affected by such abundance
differences, and in a region of each model with minimal sensitivity to these
differences. The translation to a depth of the
convective envelope, however, depends on the opacity at the bottom of the
convection zone which has major contributions from oxygen \citep{OP05}.

\subsection{Some calibrations against stellar observations}
\label{sect:OtherCals}

There are several semi-empirical calibrations of the mixing length, based on
stellar evolution calculations of binaries or stellar clusters, solving for a
common age of the stars under the observational constraints. These cannot be
compared directly with each other or with our work, since they will depend on
details of the adopted convection formulation and treatment of the outer
boundary condition, as discussed in Sect.\ \ref{sect:StelStruc}. It is,
fortunately, a widespread practice to also provide $\alpha_\odot$ of a solar
model calibrated to the present radius and luminosity \citep{gough:MLT-calibr},
and the differential behaviour with respect to the Sun should be much more
robust. In the following, we therefore compare with our results scaled to the
$\alpha_\odot$ of each study, as well as with our unscaled results.

In such a semi-empirical calibration of the
$\alpha$\,Cen system, \citet{morel:alfaCenA+B} found $\alpha_{A,B}=
($1.86\ppmm{+0.09}{-0.06},1.97\ppmm{+0.13}{-0.15}),
whereas we find values for the two components of $1.75\pm 0.03$ and
$1.76\pm 0.02$, respectively (No.\ 24 and 34 in Table~\ref{starlist} and
Fig.~\ref{cwSmax}). \citet{yildiz:aCenAB-alpha} found that $\alpha$ values of 1.64
for the primary and 1.58 for the secondary can reproduce both classic and
seismic observational constraints. These are all shown in the right-hand-side
of Fig.~\ref{obs_alfa}d. We notice that the results of \citet{morel:alfaCenA+B}
decrease with $T_{\rm eff}$, ours are both indistinguishable from the solar
value, and those of \citet{yildiz:aCenAB-alpha} increase with $T_{\rm eff}$. The
latter have no error bars, but we have assigned a $\sigma_\alpha=0.05$ based
on the scatter in $\alpha$ between the models used for his analysis.
The absolute values of different mixing-length calibrations are not
expected to agree, since they depend on details in the modelling of the
atmospheres.
The differences between the two components should be
more reliable, though, and we note that all three calibrations can be brought to
overlap when each pair is allowed to shift vertically. Also note, however, that
only the results of \citet{morel:alfaCenA+B} overlap with their
solar-calibrated $\alpha_\odot$ (same value for those two calibrations) shown
with black dotted line.

%

\citet{stassun:OriEclipsBin} performed a calibration on a pre-main-sequence
(PMS) eclipsing binary in Orion, V1174\,Ori. Their observations and analysis
favour inefficient convection of about $\alpha=1.0$ (compared with their
$\alpha_\odot=1.9$) in the pair. This is also supported by the only
0.2-0.3\,dex depletion of lithium in the 1.01\,M$_\odot$ primary component,
while the 0.73\,M$_\odot$ secondary displays more than 1\,dex depletion. These
$\alpha$ values do not agree with our calibration, however, where the primary
should have $\alpha_{\rm A}=1.06\alpha_\odot$. The secondary lies 600\,K
outside our grid, along a steep gradient, and we attempt no extrapolation.
The nearest point in our grid has $\alpha=2.02$.
\citet{stassun:OriEclipsBin} kept $\alpha$ identical for the two components,
which might account for some of their problems fitting both stars to the same
PMS evolution calculations.

\mfigur{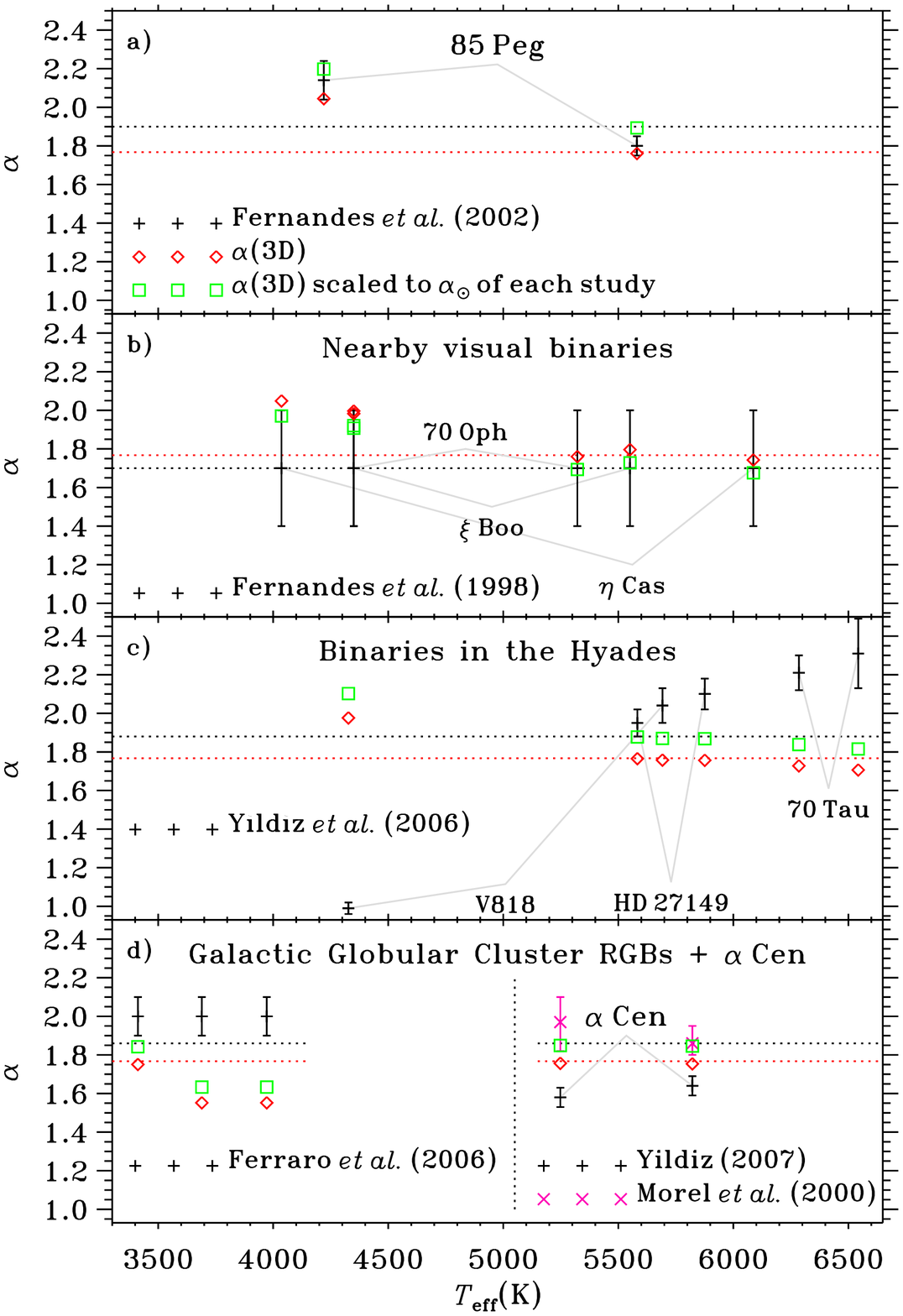}{12.4cm}
    {Comparison between our $\alpha_{\rm MLT}$ calibration (red diamonds, 
    error-bars are similar to the symbol size)
    and some semi-empirical calibrations from the literature (black
    crosses with error-bars).
    The respective solar calibrations of those
    studies are shown with horizontal black dotted lines, and the green
    squares show our results scaled to those solar values.
    The horizontal red dotted lines show our $\alpha_\odot$.
    Binaries are connected with grey lines and labelled.
    See text for details on the six semi-empirical calibrations shown here.
    Our solar metallicity calibration has been extrapolated in $T_{\rm eff}$
    to obtain the values for the metal-poor globular clusters in the left-hand
    side of panel d), as explained in the text.}

By matching observed and model properties \citet{fernandes:FundStelParmBins}
found that three sets of nearby visual binaries could all be described with a
solar mixing length of $1.7\pm 0.3$. Within the error-bars, this is consistent
with our calibration, although we would expect larger values for the cooler
secondary components, as shown in Fig.~\ref{obs_alfa}b.
\citet{fernandes:FundParms85Peg} studied the visual binary 85\,Peg and found 
$\alpha^{\rm obs}_{\rm A}({\rm 85\,Peg})=1.80\pm0.05$ and 
$\alpha^{\rm obs}_{\rm B}({\rm 85\,Peg})=2.14\pm0.10$, compared with their solar
value of 1.9. This last result is in excellent absolute agreement with ours,
$\alpha^{\rm 3D}_{\rm A}({\rm 85\,Peg})=1.76\pm 0.02$ and
$\alpha^{\rm 3D}_{\rm B}({\rm 85\,Peg})=2.05\pm 0.03$, as shown in Fig.~\ref{obs_alfa}a.

\citet{chieffi:alfa-cal} calibrated $\alpha$ against Galactic globular
clusters, by either calibrating the temperatures of the red-giant branches
(RGBs) or the slope of the main sequences. This latter method depends on
$\alpha$ being constant in $T_{\rm eff}$ and $\log g$, from the faint end
of the main sequence to 2\,mags below the turn off.
They divided their results into two
metallicity groups, $[$Fe/H$]=-1.3$ and $-2.3$, both with statistically
insignificant differences between the main sequence and the red-giant branch,
but a tentative increase with metallicity. The lack of change between the MS
and the RGB, along with the location of $t\ga 9$\,Gyr isochrones, agrees with
the triangular plateau we find in our calibration for \emph{solar metallicity}.
Their results of
$\alpha_\odot=2.25$, $\alpha_{\rm RGB}({\rm[Fe/H]}=-1.3)=1.91\pm 0.09$ and
$\alpha_{\rm RGB}({\rm[Fe/H]}=-2.3)=1.55\pm 0.23$ is consistent with a linear
increase with metallicity, assuming similar behaviour with $T_{\rm eff}$ and
$\log g$ at all metallicities. 

Fitting to observed RGBs of 28 Galactic globular clusters as
function of metallicity in the range $-2.15<[$Fe/H$]<-0.2$,
\citet{ferraro:alfaRedGiant} found that the Sun and
the RGB could be modelled with a common mixing length of 2.17, independent of metallicity. This followed
from using the classic \citet{AG89}-abundances with $(Z/X)_\odot=0.0275$.
Using instead the more modern abundances of \citet{lodders:2003}, with
$(Z/X)_\odot=0.0117$, they found $\alpha_\odot=1.86$ and
$\alpha_{\rm RGB}\simeq 2$, likewise independent of metallicity, as shown in Fig.~\ref{obs_alfa}d. This result contradicts the findings of
\citet{chieffi:alfa-cal} mentioned above.
Our results for this case are extrapolations to both lower
$T_{\rm eff}$ and $\log g$, so they should be interpreted with caution. This
particular extrapolation almost follows the contours of
$\alpha(T_{\rm eff}, \log g)$, though, making the extrapolation less suspect.
The extrapolation in metallicity is potentially a bigger issue, but the
apparent lack of metallicity dependence of their results, warrants a comparison
with our [Fe/H]=0.0 grid.

\citet{piau:RGalpha+R} studied the red (cool) edge of the RGB, based on a
sample of 38 nearby Galactic disk sub-giants and giants with interferometrically
determined radii.
They find that the red edge, constrained by six stars, is best fitted with
$\alpha=1.68$ compared with their solar calibration of $\alpha_\odot=1.98$.
They also find that a single value fits the observations over a decade of
luminosity, in agreement with our calibration showing RGB evolution along
contours of $\alpha$. Our results suggest, however, that $\alpha_{\rm RGB}$,
for their mass of 0.95\,M$_\odot$, should be only 0.04 smaller than
$\alpha_\odot$.
Their sample is close to solar metallicity, with the spread of the observations
around the evolution tracks being accounted for by the metallicity effect on
the stellar models.

Studying binaries in the Hyades, \citet{yildiz-alfa-cal} found a mixing length
that varies a factor of 2.3 over the MS mass-range of 0.77--1.36\,M$_\odot$.
Our calibration suggests a factor 1.15 variation instead, and in the opposite
sense (See Fig.~\ref{obs_alfa}c).

With further constraints from asteroseismic observations and analysis
\citet{savita:Kpl22sols} found $\alpha$ ranging from 1.6 to 2.2 for 22
main sequence {\it Kepler} targets ranging from 0.8 to 1.3\,M$_\odot$. They found a
possible increase of $\alpha$ with $T_{\rm eff}$, possibly with a bump around
$T_{\rm eff}\sim 5\,800$\,K of amplitude $\sim$0.2, but with sizable scatter. 
\citet{bonaca:KeplerAlpha} similarly found a weak increase with $T_{\rm eff}$
(without a bump, though) and also with $\log g$ and metallicity, [Fe/H],
for the main-sequence stars in their sample. Both of these results are
contingent on the prescriptions for accounting for the \emph{surface effect}:
the systematic frequency shift between observations and 1D MLT models that
signal differences around the upper turning point of the modes
\citep{jcd:solar-freq-shifts,hans:SeismNearSurf}. The surface effect
constitutes a sizable systematic effect in both analyses. Recently
\citet{metcalfe:AMP42Kepler} analysed 42 {\it Kepler} targets including
solar-like oscillators, F-stars and subgiants, taking special care to
physically constrain the surface effect. They find $\alpha$-values that
agree with our calibrations, although the dependency on both $T_{\rm eff}$ and
$\log g$ is stronger than ours. The metallicities of their sample ranges from
[Fe/H]$=-0.6$ to +0.5 with an average and RMS-scatter of $-0.06\pm 0.20$, as
well as a single metal-poor star of [Fe/H]$=-1.14$. They find that $\alpha$
increases with [Fe/H], and that the behaviour with $T_{\rm eff}$ and $\log g$
does not change significantly when only the solar metallicity,
$|[{\rm Fe/H}]| < 0.2$, sub-sample is analysed.

In general, the range of $\alpha$ values from our calibration is smaller than
what is suggested by the various types of stellar model fitting employed above.
This can be due to a number of issues, both in the modelling and in the
interpretations of observations. In the former category, any inadequacies or
inconsistencies in the treatment of atmospheric opacities and {\Ttau}s
will be absorbed into the calibrated values of $\alpha$.
In the latter category, the translation between observations (whether
photometric or spectroscopic) and $T_{\rm eff}$
is one of the most important. To take the temperature of a star
is a non-trivial endeavour, but recent advances have been made by, e.g.,
\citet{casagrande:IRfluxTeff} and \citet{melendez:SunTwin-uvbyPhot} based on
observations of solar twins (stars that are spectroscopically and
photometrically indistinguishable from the Sun), and by
\citet{huber:FundParamsKeplerCHARA} based on asteroseismic analysis of {\it Kepler}
and CoRoT observations, coupled with interferometric measurements of stellar
radii. These calibrations of the $T_{\rm eff}$-scale are being used in an
absolute calibration of a range of photometric systems, based
on synthetic photometry of the grid of simulations that we present here.

The mixing length of stellar atmosphere models has also been calibrated against
stellar spectra, which is a profoundly different kind of calibration.
\citet{fuhrmann:Balmer-lines}, \citet{vant-veer-menneret:BalmerTeff} and
\citet{gardiner:BalmerTeffAFG} all found that only a small value of $\alpha=0.5$
can reproduce the shape of solar Balmer lines. This is half of the atmospheric
mass mixing length found by \citet{trampeda:mixlength}, and thus does not
reflect the scale of mixing. Nor does it correspond to the entropy jump in the
atmosphere, which is effectively the quantity calibrated when observed global
properties are fitted with evolution models. Instead it reflects the larger
super-adiabatic gradient of the warm granules, which dominate the emergent
spectrum, including the Balmer lines, due to their brightness
\citep{asplund:3DabundAnalysis,trampedach:Rome2009}. The solar Balmer lines have
been successfully modelled in NLTE, based on a 3D convection simulation
\citep{tiago:suneHlinesLimbdark}, and less successfully with 1D
PHOENIX models \citep[using $\alpha=1.0$]{hauschildt:nextgen-atm} and
MARCS models \citep[using $\alpha=1.5$]{MARCS-2}.

The effects on stellar evolution models, of varying both the {\Ttau} and
$\alpha$ according to our simulations, will be addressed in Paper\,III.

\section{Conclusion}
\label{sect:concl}

We have calibrated the MLT parameter $\alpha$ by matching 1D envelope models
with 3D RHD simulations, and established a significant variation of $\alpha$
with stellar atmospheric parameters $T_{\rm eff}$ and $g_{\rm surf}$.
Our results show a triangular plateau with $\alpha\simeq 1.76$ stretching from the
bottom of the red-giant branch at $\log g=3.3$ to $T_{\rm eff}\sim 6\,400$--4\,800\,K on the main sequence. A similar plateau was found in the
calibration by LFS against 2D simulations. This plateau includes
the Sun, as well as the $\alpha$\,Cen system and their evolution so far. This
suggests a common and constant $\alpha$ for the evolution from the main sequence
of these three stars, but much of the pre-main-sequence evolution would have
occurred with higher $\alpha$, as shown in Fig.~\ref{cwalfa}.

As stars ascend the giant branch we see that they evolve largely along
contours of constant $\alpha$. During this evolutionary stage $\alpha$
decreases with mass, from $\alpha(M=0.4)=1.75$ to
$\alpha(M=4.8\,$M$_\odot)=1.56$
(simulations No.\ 2 to 3). The largest gradients in $\alpha$
occur during the evolution of low mass stars, $M<0.8\,$M$_\odot$, and for
higher mass stars, $M>1.2\,$M$_\odot$, crossing the Hertzsprung gap after the
turn-off from the main sequence.

Although various values of $\alpha$ have been considered in the modelling of
stellar evolution, an $\alpha$ varying during the evolution of a star has, to
our knowledge, not been tried yet. Results of such evolution calculations
are presented in Paper\,III.

In Sect.~\ref{sect:StelStruc} we investigated how changes to the radiative part of
the outer boundary affect the structure of a star, using the depth of the outer
convection zone as a global measure. We evaluated the linear response of the
change in depth of the convection zone caused by changes in atmospheric
opacity, {\Ttau} and mixing length, respectively. Our analysis in
Sect.~\ref{sect:dCZ} shows that the convection zone is about equally sensitive to
the three kinds of changes, and consequently different MLT parameter triplets
can easily result in the same global properties of a stellar model. References
to a particular mixing length are therefore less useful unless accompanied by
references to the atmospheric opacity and {\Ttau}.

We also compared the effects of various commonly used assumptions about the
{\Ttau}, and concluded that using the old solar {\Ttau} as given by
\citet{krishna-swamy:lines-K-dwarfs}, scaled to the $T_{\rm eff}$ of each star,
results in convection zones that are shallower by up to 7\% of the radius
(see Fig.~\ref{cwdqKS}). Using scaled
versions of the {\Ttau} from the solar simulation also introduces systematic
effects, causing deeper convection zones in stars cooler than the Sun and
shallower ones in warmer stars. Using a 5000\,{\AA} {\Ttau} with a Rosseland
opacity has a similar but smaller effect. We recommend a consistent usage of
\Ttau{s} and their corresponding opacities in stellar structure and evolution
calculations.

We stress that the choice of $\alpha$ depends on the choice of atmospheric
physics, {\ie}, {\Ttau} and atmospheric opacity. Employing the commonly used
scaled solar {\Ttau} will alter the effect of $\alpha$, as shown in Sect.\
\ref{sect:Ttaueff} and Sect.\ \ref{sect:constTtau-alpha}. We recommend that
our calibrated $\alpha$ values be used with the atmospheric opacities and
individual {\Ttau}s from Paper\,I.

As ground-based and especially space-borne asteroseismology with CoRoT and
{\it Kepler} is now
providing strong constraints on the structure of stars other than the Sun,
stronger demands are placed on our theoretical models. Keeping our models
ahead of the asteroseismic capabilities of the next missions, TESS and
PLATO, and the recently launched astrometry mission, Gaia, is a great challenge for the
modelling community.

An absolute calibration of the mixing-length parameter, $\alpha$, is the
first step towards improving the treatment of convection in stellar structure
models. A fundamentally improved formulation of convection is of course
desirable, but has proven rather difficult to come by. Various attempts
have been made to rectify this situation. \citet{canuto:R-stress1} present
a formulation based on fully developed turbulence, which, however, does not
account for the steep density gradient and the inherent asymmetry between
up- and downflows. \citet{fox:formconv1} base their model on 3D hydrodynamical
simulations of convection, and this is probably the most promising way forward.
A number of approximations render their results less than optimal for the
next generation of convection models, however.

With the connection between MLT and realistic 3D convection simulations,
discussed in Sect.\ \ref{sect:MLT-3D} and by \citet{trampeda:mixlength}, we
find a properly calibrated mixing-length formulation, with the
mixing length being proportional to the pressure scale height, to be the best
choice for the time being.

\paragraph*{{\bf Data Retrieval:}} A file with the calibrated mixing length
\label{sect:download}
parameters and Fortran\,77 routines for reading and interpolating the data can
be downloaded from: \url{http://cdsarc.u-strasbg.fr/viz-bin/qcat?J/MNRAS/442/805} The data-file contains both the
radiative Hopf functions, $q(\tau_{\rm Ross})$, as found in Paper I, and the
calibrated mixing-length parameter, $\alpha$, as function of atmospheric
parameters, $T_{\rm eff}$ and $\log g$. The URL also contains the routines
necessary for setting up and interpolating in the triangulation of the
irregular grid of simulations \citep{renka:triangulation}. Finally, we also
supply a simple user-level function to include in stellar structure codes,
which does not require any knowledge of the data or the details of the
triangulation.

The {\tt OPINT} opacity interpolation package can be downloaded from
\url{http://phys.au.dk/~hg62/OPINT}, together with the atmospheric
opacities from our calculation (Paper\,I, Sect.\ 3.1), merged with interior
OP opacities (cf.\ Sect.\ \ref{sect:env}).

\section*{Acknowledgements}
We thank the anonymous referee for helpful comments which have substantially
improved the presentation.
We are grateful to Werner D\"appen for granting us access to the MHD-EOS code
and to R.\ F.\ Kurucz for providing us with his
tables of opacity distribution functions.
Funding for the Stellar Astrophysics Centre is provided by The Danish National Research Foundation (Grant DNRF106). The research is supported by the ASTERISK project (ASTERoseismic Investigations with SONG and {\it Kepler}) funded by the European Research Council (Grant agreement no.: 267864).
We would like to thank the Australian Partnership for Advanced Computations (APAC) for generous amounts of computer time.
RT acknowledges funding from NASA grants NNX08AI57G and NNX11AJ36G and from
the Australian Research Council (grants DP\,0342613 and DP\,0558836)
RFS acknowledges NSF grant AGS-1141921 and NASA grant and NNX12AH49G.
This research has made use of NASA's Astrophysics Data System Bibliographic Services.


\bsp

\label{lastpage}

\end{document}